\documentclass[twocolumn]{aa} 
 
\usepackage{natbib}      
\bibpunct{(}{)}{;}{a}{}{,} 

\usepackage{txfonts}
\usepackage{caption}
\usepackage{subcaption}
\usepackage{graphicx}
\usepackage{bm}
\usepackage[utf8]{inputenc}
\usepackage{float}
\usepackage{stfloats}
\usepackage{color,soul}
\usepackage{xcolor}

\usepackage{hyperref}
\hypersetup{
colorlinks=true,
linkcolor=blue,
urlcolor=blue,
citecolor=blue,
}
\begin{document} 

\newcommand{\twolin}{\leavevmode \\ \leavevmode \\}
\newcommand{\trelin}{\leavevmode \\ \leavevmode \\ \leavevmode \\}
\newcommand{\graphics}{plots_finalfinal}
\newcommand{\graphicsmore}{plots_finalfinal}
\newcommand{\graphicsmoree}{plots_finalfinal}

\newcommand{\gf}{plots_finalfinal}
\newcommand{\gfstr}{plots_finalfinal/streams}
\newcommand{\cmg}{\rm\, cm^2 \, g^{-1}} 
\newcommand{\gcm}{\rm\, g \, cm^{-3}} 

   \title{The influence of dust growth on the observational properties of circumplanetary discs}
	 \titlerunning{Spectra of CPDs}
	
   \author{Matth\"aus Schulik,
          \inst{1,2}
          Bertram Bitsch
          \inst{3}, 
          Anders Johansen
          \inst{2,4}
           and
		    Michiel Lambrechts \inst{2,4}}
			
	\authorrunning{Schulik et al.}

   \institute{
   Imperial College London, Kensington Lane, London W1J 0BQ, UK\\\email{mschulik@ic.ac.uk}
   \and
   Lund Observatory, Box 43, S\"olvegatan 27, SE-22100 Lund, Sweden
    \and
    University College Cork, College Rd, Cork T12 K8AF, Ireland
    \and
    Center for Star and Planet Formation, Globe Institute, University of Copenhagen, Øster Voldgade 5–7, 1350 Copenhagen,
Denmark
    }

   \date{Received ...}
 
\abstract{Dust growth is often indirectly inferred observationally in star-forming environments,
theoretically predicted to produce mm-sized particles in circumstellar discs, and also presumably witnessed by the predecessors of the terrestrial meteoritic record. For those reasons it is believed that young gas giants under formation in protoplanetary discs with putative circumplanetary discs (CPDs) surrounding them, such as PDS 70c, should be containing mm-sized particles. We model the spectra of a set of CPDs, which we obtained from radiation hydrodynamic simulations at varying Rosseland opacities $\kappa_{\rm R}$.
The $\kappa_{\rm R}$ from the hydrodynamic simulations are matched with consistent opacity sets of  {ISM-like composition, but grown to larger sizes}.  {Our} high $\kappa_{\rm R}$ hydro data nominally corresponds to 10 $\mu$m-sized particles, and  {our} low $\kappa_{\rm R}$-cases correspond to mm-sized particles. 
 {We investigate the resulting broad spectral features at first while keeping the overall optical depth in the planetary envelope constant.} 
Dust growth to size distributions dominated by millimeter particles generally results in broad, featureless spectra with black-body like slopes in the far-infrared, while size distributions dominated by small dust develop steeper slopes in the far-infrared and maintain some features stemming from individual minerals.
We find that significant dust growth from microns to millimeters can explain the broad features of the PDS 70c data, when upscaling the dust masses from our simulations by $\times$100. Furthermore our results indicate that the spectral range of 30-500 $\mu$m is an ideal hunting ground for broadband features arising from the CPD, but that longer wavelengths observed with ALMA can also be used for massive circumplanetary discs.}



   \maketitle
%

\section{Introduction}

Planet formation is a by-product of the star formation process. A significant mass fraction of the collapsing cloud that makes up a future star \citep{Hueso2005, pascucci2016, manara2018, tychoniec2018} collapses into a rotationally supported circumstellar disc (CSD) which forms planets \citep{najita2014}. The dust component that is incorporated into the CSD will initally bear resemblance to dust found in the interstellar medium (ISM). This ISM dust consists of amorphous olivine mineral and amorphous graphite components with a maximum size of $\sim$0.3 $\mu$m \citep{draine2003}, whereas the nature of the carbon-bearing species could as well be organics grown from PAHs \citep{pollack1994}. Beginning in the ISM, dust plays an important double-role throughout the entirety of the planet format process. On one hand it is providing the mass for the larger building blocks of planets, on the other hand it is responsible for cooling and compactifying the gaseous envelopes of the forming planets.

The earliest stages of planet formation can be accessed via studies of the spectra of nearby star-forming regions in the Milky Way \citep{martin2012} and prestellar cores \citep{chacon2019}. Those works make it clear that dust undergoes a significant, and environment-dependent, evolution as it passes through ever denser mass reservoirs. Indirect evidence from class 0 \citep{harsano2018} and class II discs around brown dwarves \citep{pinilla2017} and detailed dust coagulation models \citep{birnstiel2012} furthermore suggest that the dust can grow to millimeters in size in a number of sources, although it remains unclear whether the growth occurs in the olivine, organics or an icy component. Dust then can continue to form planetesimals, terrestrial planets and nucleate the cores of giant planets \citep{lambrechts2014, morbidelli2015}. 

Cores of giant planets surround themselves with a significant amount of gas, called the protoplanetary envelope. This envelope can be thought of as an enourmous extension of the protoplanetary atmosphere, which can still be physically connected to the parent CSD, particularly before gap-opening. Because the gas is unable to form a significantly overlapping spectral line continuum in the low-pressure environment of envelopes, the envelope cooling capability is dominated by the dust continuum opacities. Those opacities have to be low enough in order to allow proto-giant envelopes to radiate their compressional heat and reach runaway gas accretion before the CSD dissipates \citep{pollack1996, pisoyoudin2014}.

The opacity of a dust mass is inversely proportional to the size of the individual particles, because larger dust distributes the same mass in a lower effective surface area and is therefore generally less opaque \citep{ossenkopf1994}. As a consequence, ISM-like dust has a too high opacity to cool gaseous envelopes in time to form giants\citep{hubicky2005}. However, it was shown that the envelope provides the incorporated dust with the dense environments necessary to grow up to mm-sizes and hence, the required low opacities to form gas giants can be reached self-consistently \citep{movshovitz2010, mordasini2014a}. 
The results from the latter works indicate that the key cooling parameter, the Rosseland opacity, should be on the order of $\kappa_{\rm R}=10^{-3}-1\, \rm cm^2 \, g^{-1}$, corresponding to single-grain sizes of cm to 10 $\mu$m,  {with variations around this number allowed}. 

Computational models in 2-D and 3-D predict the envelopes of giant planets to form structures, in the form of circumplanetary discs (CPDs) \citep{lubow1999, kley1999, ayliffe2009b, tanigawa2012, judith2016, maeda2022, li2023}, when simplified assumptions about thermodynamics, and particular the cooling rates are made. The CPD would furthermore form an environment for moon formation \citep{canup2006, judith2016, shibaike2019, ronnet2020, anderson2021}. We recently were able to show in \cite{schulik2020} that the formation of a CPD is possible at $\kappa_{\rm R}=10^{-2}\, \rm cm^2 \, g^{-1}$, while it is suppressed at $\kappa_{\rm R}=1\, \rm cm^2 \, g^{-1}$. Both our simulated giant planet envelopes accrete gas vigorously, but only for the low opacity does the accreted gas pass through the CPD in the midplane, leading to mass build-up in the CPD. 
 {The latter result has recently been understood to belong to a general class of cooling criteria \citep{krapp2024}, while viscosity-based criteria are also being investigated \citep{lega2024}.}

We now know of two planetary objects, PDS 70b and PDS 70c, which have strong evidence for them to being young, actively accreting protoplanets in the PDS 70 system  {\citep{Keppler2018, haffert2019, aoyama2019, zhou2021}}.  {Recently, a third object, PDS 70d of yet unknown nature, has been proposed \citep{christiaens2024}}. This system, having an age of 3-5 Myrs and a possible CPD around PDS 70c \citep{isella2019, benisty2021}, shall be used as an inspirational case for this work.  {Further analysis of the system was carried out in \cite{bae2019}, who found that the two planets are close to a 2:1 resonance and the dust ring widths are consistent with filtration by the planets, emphasizing this systems potential as a laboratory for  planet formation theory}.  {We investigate the circumplanetary discs of simulated Jupiter-mass planets, a mass compatible with the extremes of the mass-determination from \citet{wang2020}.} We emphasize the notion of a 'protoplanet' as a planet that is observed during its ongoing formation process, and it is the latter that makes the PDS 70 planets interesting. Particularly the finding of a beam-resolved sub-mm source in PDS 70c now opens the possibility of studying the building material of protoplanets and CPDs up close, but this requires an understanding of the dust that provides the PDS 70c sub-mm emission. 

With the understanding that CPDs need low opacities to form, the next step is to investigate possible dust components that are consistent with those values, and whether one could infer the presence of those components from spectral observations such as those of PDS 70c. Additionally, it is of importance to ascertain the source of the dust emission flux, as there is a possibility that the observed spectrum of a protoplanet is a mix of planet and CPD contributions \citep{haffert2019,judith2019}.

Therefore in this work, we use the density-temperature structures from the 3D radiation hydrodynamics simulations, presented initially in \cite{schulik2020}. We then post-processed those simulations with a number of plausible dust populations that are consistent with low $\kappa_{\rm R}$, in order to generate synthetic spectra of circumplanetary discs and assess their importance relative to the planets. Our results indicate that if mm-sized particles dominate the particle distribution, those will act to significantly grey the spectrum of planet + CPD. Furthermore, spectra of objects that emit at a number of temperatures are always a sum of black-bodies. Therefore, depending on the exact geometry of emitting source temperatures, this summation can leave traces by flattening the far-infrared slope of a spectrum. This would be the analogous process like in the spectra of circumstellar discs, which we see weak evidence of in our synthetic observations. Finally, we find that the contribution of the circumplanetary disc to the total spectrum generally peaks between 30 $\mu$m and 500 $\mu$m. 
 {Our work, while not attempting to fit an exact model to PDS70c, makes it therefore nonetheless plausible that the broadness of said spectrum indicates dust growth.}

This paper is organized as follows. In Section 2, we introduce briefly the density-temperature structures of our CPD simulations. Furthermore, we outline the provenance of the laboratory dust analogues which we use and discuss size, fluffines, composition and ice coating parameters of the basic dust particles employed.
The same section discusses the computational tools employed to arrive at the final data product, which are images and spectra ranging from the NIR to the submm. In Section 3, we briefly outline the geometry of the emission surfaces for one example CPD at various wavelengths, and present the general analysis procedure. We later employ all dust opacity carriers of interest to discuss synthetic spectra and imaging. Section 4 summarizes our findings and discussed further needed work.

\begin{table}
\centering
\caption{Set of hydro simulations and matching grain mixes for post-processing }
\label{tab:sim_data}
\begin{tabular}{c c c c c}        
\hline\hline                 
Sim$\#$ & $\kappa_R$-hydro $\rm [cm^2/g]$ & $f_{\rm DG}$ & grain mix & Comment \\    
\hline
1 & 1 & 1 & S & Early\\
\hline
2 & 0.01 & 1 & L & Early\\
3 & 0.01 & 0.01 & S & Early \\
4 & 0.01 & 0.001 & ISM & Early \\
\hline
5 & 0.01 & 100 & L & Late\\
6 & 0.01 & 1 & S & Late\\
7 & 0.01 & 0.1 & ISM & Late\\
\hline
8 & 0.01 & 0.1 & ISM & Late, cold \\
9 & 0.01 & 100 & L & Late, cold \\
\hline
	\end{tabular}
	\\[5pt]
	\caption*{Notes on columns. In this work, we create post-processed observations of radiation hydrodynamic data generated with values of $\kappa_R$-hydro and match those average opacities with a mix of wavelength-resolved opacities of more realistic dust mixes. To reach average low opacities, one can employ the nominal mix of large particles, or smaller particles, with the dust-to-gas $f_{\rm DG}$ ratio adjusted downwards, which is done for sims 2-4. Every mix is a scaled \cite{draine2003}-distribution, and has a certain major particle size $a_{\rm major}$ characterizing it. The large 'L' mix has $a_{\rm major}=1$ mm, the small 'S' mix $a_{\rm major}=10$ $\mu$m and the 'ISM' mix $a_{\rm major}=1\, \mu$m. 'Early' type simulations denote nominal densities. Higher dust-to-gas ratios are used in sims 5-9 to mimic dusty discs 100 times more massive than actually given in the simulations, called 'late' simulations, as they shall represent a stage of late CPD accretion. Sims 8 and 9 employ additionally a temperature reduction to mimic late-stage cooling of the disc. }
\end{table}

\section{Methods}

In order to generate synthetic spectra and observations, we use the radiative transfer code \textit{radmc3d} \citep{dullemond2012} v0.41 \footnote{\url{http://www.ita.uni-heidelberg.de/~dullemond/software/radmc-3d/}}. This code needs density data, temperature data and wavelength-resolved opacities for all dust species involved. The latter need to be computed with separate tools for individual dust parameters from the laboratory optical constants $n$ and $k$.
In the following paragraphs we describe our choices of each of those components.
Table \ref{tab:sim_data} summarizes the combinations of hydrodynamic data and dust post-processing parameters used.

\subsection{Optical constants, dust parameters and \text{radiation transport} tools}

Dust populations in the interstellar medium consist of $\sim$ 0.3 $\mu$m amorphous olivines and amorphous carbonates, with additional smaller components of amorphous carbonates at particle sizes $a_P \approx 10$ nm and PAHs of $\approx 2$ nm sizes \citep{draine2003, compiegne2011, planck2014}. The PAH macromolecules however do not seem to be important for the spectra of protostellar discs and are absent in the presolar meteoritic record \citep{scott2007} which is why we neglect this component in our work.

Moreso, interstellar dust does not seem to be a good proxy for planetary building blocks. The terrestrial meteoritic record shows that chondrules of up to cm sizes formed in the solar system continuously for 3 Myrs after CAIs \citep{connely2012}, and are mostly products of repeated heating and crystallisation in the early solar system \citep{scott2007}. While their high-pressure, high-temperature processing is thought to have happend in extreme environments, such as nebular shocks of various kinds, their mass assembly must have happened prior to the thermal processing and hence implies a presumably fluffy, precursor population of dust particles similar to what is seen in cometary (protosolar) dust \citep{brownlee2014, guttler2019}, only larger, at mm-to-cm sizes. Together with the previously mentioned indirect evidence for mm-sized particles in CPDs \citep{pinilla2017, harsano2018} we think it is reasonable to assume that a population of mm-sized particles would dominate the dust mass budget in a several Myr old source orbiting a fully formed giant planet, like PDS 70c.

The exact number distribution and chemical makeup of those grown particle populations will be of paramount importance for spectral modeling. Motivated by analytical models \citep{birnstiel2012} and observations of the ISM \citep{draine2003, compiegne2011} we use three mixes of dust distributions in order to match the two different average Rosseland opacities from our hydrodynamic simualtions. The average Rosseland opacity $\kappa_{\rm R}$ is defined by 
\begin{align}
\kappa_{\rm R} = \left( 
\frac{\int_0^{\infty} d\lambda \; (\kappa(\lambda))^{-1} \; \frac{\partial B}{\partial T} }
{\int_0^{\infty} d\lambda \; \frac{\partial B}{\partial T}}
\right)^{-1},
\label{eq:mean_opacities}
\end{align}
where $\kappa_{\rm \lambda}= \sum_i \kappa_{i}(\lambda)$ are the non-averaged opacities with contributions from all dust species and $B=B(\lambda,T)$ is the Planck function for a black-body of temperature $T$. As we use two simulations with hydrodynamic data for constants $\kappa_{\rm R} = (1, 0.01)\rm\,cm^2\,g^{-1}$, those values must be matched at least approximately by the dust populations in order to ensure self-consistent modelling.

\begin{figure*} 

 \hspace*{+0.0cm}
  \begin{subfigure}{0.475\textwidth} 
	\centering
	\includegraphics[width=1.0\textwidth]{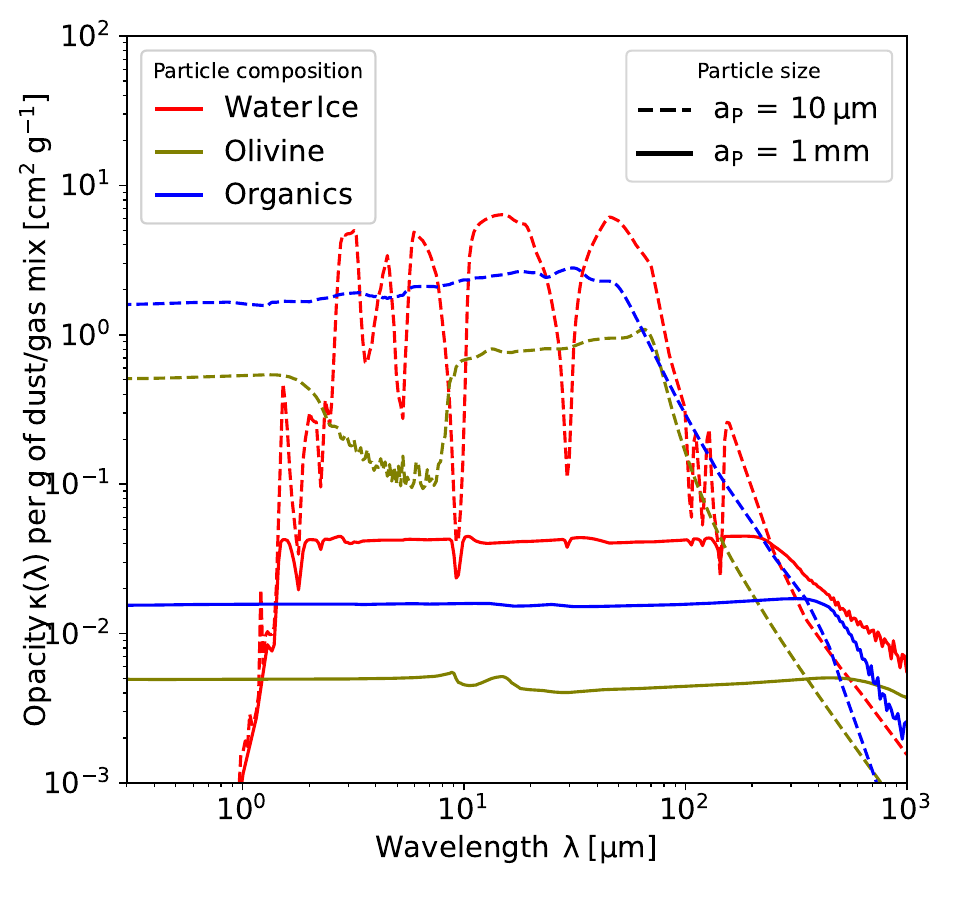}
   \end{subfigure}%
    \hspace*{+0.5cm}	
	  \begin{subfigure}{0.475\textwidth} 
	\centering
	\includegraphics[width=1.0\textwidth]{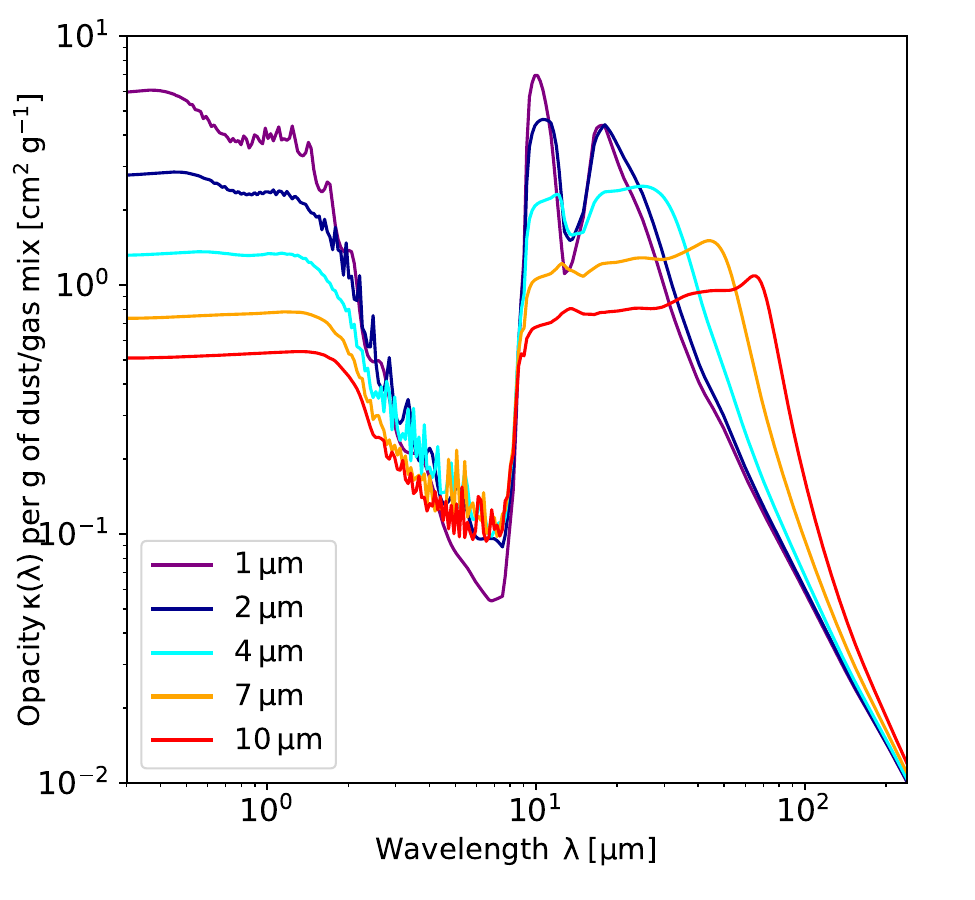}
   \end{subfigure} 
   
   \caption{Computed opacity functions for water ice, olivine and organics for individual components of single particle sizes (\textit{Left}) {, scaled according to the available dust mass according to Eqn. \ref{eq:dust_densities}.} The mm-sized particles make up the 'L' particle mix, whereas the 10 $\mu$m particles make up the 'S' particle mix (see also in the main text and Table \ref{tab:sim_data}). In general it becomes evident that 10 $\mu$m particles are already fairly grey, except for the water ice. Furthermore, we show how this greyness, and particularly the disappearance of the prominent 10 $\mu$m feature for Olivine, results from step-wise dust growth (\textit{Right}). The opacity functions for the solid lines on the right are computed for  {pure Olivine} at single sizes only. The step-wise increase in size highlights the disappearance of the 10 $\mu$m-SiO feature in our spectra for dust sizes $\geq$$10\,\mu$m, as then the size parameter $2\pi a/\lambda$ enters the interference regime of Mie-scattering.
   }
\label{fig:opacities}
\end{figure*}

In order to compute the  $\kappa_i(\lambda)$ one needs the real and imgainary optical constants, $n(\lambda)$ and $k(\lambda)$, of the material in question. We use amorphous olivine, organics and water ice with the optical constants taken from the St. Petersburg-Jena-Heidelberg database of optical constants \citep{henning1999, jaeger2003} (the 'new' dataset\footnote{\url{http://www2.mpia-hd.mpg.de/home/henning/Dust_opacities/Opacities/opacities.html}} with the 'normal' olivines\footnote{\url{http://www2.mpia-hd.mpg.de/home/henning/Dust_opacities/Opacities/RI/new_ri.html}}). We convert the optical constants into $\kappa_i(\lambda)$ in the simplest case of spherical grains by specifying particle size $a_P$ and particle density $\rho_P$ and computing their Mie-scattering properties with the \cite{bohren1983} code.

The resulting opacity functions for the two key dust sizes in this work, are computed by the \cite{bohren1983} code and can be seen in Fig. \ref{fig:opacities} (\textit{Left}). The mm-sized grains display very grey spectral behaviour throughout the entire wavelength range, hence we can expect the total spectrum of the CPD to be a convolution of black bodies. The small dust is nevertheless also fairly grey, as can be very clearly seen in the absence of the otherwise notable 10 $\mu$m feature for the olivine dust component. In Fig. \ref{fig:opacities} (\textit{Right}) we show that this disappearance of the 10 $\mu$m feature is due to a dramatic transition of spectral flattening that happens between 1-10 $\mu$m of particle size, as would be expected from Mie theory in the interference regime $2 \pi a\approx \lambda$.

The dust mixes we use are constructed as follows. Each dust mix consists of amorphous olivine and organics that follow a scaled \cite{draine2003}-like size distribution. This distribution consists of several sub-distributions. The sub-distribution representing the largest particles is between 0.1 and 1 times the maximum size, with a MRN-like distribution between them. We refer to the latter as the major size for a mix, $a_{\rm major}$. This major sub-distribution exists for both amorphous olivine and organics. The \cite{draine2003}-distribution contains a another sub-distribution of smaller, amorphous carbon species, for which we again use organics, that has a gaussian radius size distribution centered at $1/30$ of the major size. We refer to this component as the minor particle size $a_{\rm minor}$, which contains $1/30$ of the large dust particles mass. Finally, water ice exists exclusively as coatings on the organics and olivine grains and distributes its total mass and volume equally on their surfaces.

We employ three types of such mixes in this work, each characterized fully by the preceeding description and the statement of its $a_{\rm major}$. The large dust mix, or 'L' in Table \ref{tab:sim_data}, is has by $a_{\rm major} = 1$ mm and has been primarily designed to fit the average $\kappa_{R}=\rm 0.01\; cm^2 \, g^{-1}$ in our radiation hydrodynamic simulation. The small dust mix, or 'S', has $a_{\rm major} = 10\rm \;\mu$m, and the 'ISM' mix has maximum size of 1 $\mu$m. 

 {The total dust mass that this scaling correpsonds to is $2\times10^{-5} m_{\oplus}$, deduced from the gas mass in our nominal circumplanetary disc of $2\times10^{-3} m_{\oplus}$ with $f_{\rm DG}=0.01$. In our later simulations, runs 5-9, which emulate more mature circumplanetary discs (see also Table \ref{tab:sim_data}) both masses are increased artificially by a factor 100, by upscaling all volumetric gas densities. The increased dust masses correspond to enough solid material to build an analogue of the Galilean moons.}

We generate the dust density $\rho_{i,\rm D}$ of each dust sub-distribution $i$ from the gas density $\rho_{\rm G}$ (described below) using the dust-to-gas ratio $f_{\rm DG}$ from Table \ref{tab:sim_data}, and a species fraction $f_{i}$, so that it is
\begin{align}
     \rho_{i,\rm D} = \rho_{\rm G} \,\times\, f_{\rm DG} \,\times\, f_{i}\,\times\,1/100
     \label{eq:dust_densities}
\end{align}
 {where the index $i$ encapsulates the species, i.e. }$i \in\rm \{Ol,major; C,major; C,minor; H_2O  \}$. The numbers $f_{i}$ are taken from \cite{semenov2003} which are consistent with observations of the ISM, hence for dust of approximately $\mu$m in size, but lacking laboratory data on the relative growth speeds of dust compounds, we take those fractions for all dust mixes. Another reason to employ those mass fractions is because \cite{draine2003} do not measure water ice abundances, which we need for our modeling. This also implies that we do not prescribe any differential settling, which is consistent with the high $\alpha\approx 10^{-2}$ that was used to generate the hydro data.

The mass fractions are $f_{\rm H_2O}=0.555$, $f_{\rm Si} = 0.264$ and $f_{\rm C, \rm major} = 0.353$, and we estimate from \cite{draine2003} that $f_{\rm C, \rm major} = 30 \times f_{\rm C, \rm minor}$. Furthermore, the sublimation temperatures are $T_{\rm H_2O,\, sub}=$155 K, $T_{\rm Si,\, sub}=$1500 K and $T_{\rm C,\, sub}=$410 K. 

The optical properties of the water ice coatings in our model are ignorant of their respective substrate. This does not reflect reality \citep{ossenkopf1994}, as an electromagnetic wave penetrating into an icy dust coating will interact with its substrate, if the coating is thin enough, but we nevertheless maintain this simplificiation here. In our implementation, the ice spectrum simply replaces that of olivine and organics in regions where $T<T_{\rm sub, H_2 O}$, which is the sublimation temperature of water ice.
Hence, cold regions in which the water has not sublimated can be dominated by water ice emissions, depending on the exact model.

Computing the average Rosseland opacity now with those mixes of the specified sizes, the result would always be a factor 3 times larger than the $\kappa_{\rm R}$ from the hydro data. One could fix this by adopting even larger particles, but this would erase all spectral features. Due to this reason we adjust the dust-to-gas ratio by this mismatch factor of 3.

The overall generation of spectra with \textit{radmc3d} is performed without the use of scattering. A detailed investigation of scattering effects for a different setup has been presented in \cite{judith2019}. Their analysis concludes that scattering would eradicate features between 1-10 $\mu$m efficiently and overlay it with a scattering slope, but is naturally very weak for longer wavelengths. Because in this work, we are particularly interested in the continuum emission properties in the near-infrared to sub-mm range, we neglect scattering effects.

We treat gas opacities only in a very simplified manner, as to imitate their effect as background opacity source in very hot regions, where all dust has evaporated. To this end, we set the gas opacity function to a constant grey value $\kappa_{\rm G}(\lambda) \equiv 10^{-5}\rm\, cm^2/g$. However, for all practical purposes we find that the inclusion or exclusion of the gas in this manner has no effect on the spectra  {because the spectral signal is generally dominated by cool dust}. 
Regions where  {hotter dust and gas emission} might play a role, due to the sublimation of all dust components, are however hidden from the observer, as all $\tau=1$ surfaces in our models lie exterior the olivine sublimation front (c.f. Fig. \ref{fig:opacity_map}, dark blue line).

 { All this information is fed to the open source code \textsc{Radmc3d} \citep{dullemond2012} to generate the objects' spectra. We employ this code without a stellar irradiation source, in which the source function for the photons is purely given by the temperature field from our radiation hydrodynamic simulations. The code then solves for the radiation transport through the density and opacity fields, fed into the code as separate input. The optical depth in each cell is chosen such that the optical depth of the dust mixture in each cell corresponds to the optical depth in our radiation hydrodynamic simulations.}

We now proceed to describe the radiation hydrodynamic data and how it was processed.

\subsection{Hydrodynamic data}

In a previous work \citep{schulik2020}, we have carried out global 3D radiation hydrodynamic simulations of the gaseous envelopes around Jupiter-mass planets.  {For an overview, we shall briefly summarize some of the most important points from that work.  Our protoplanets are represented by a gravitationally smoothed potential. In units of $r_{\rm H}=5.2\, \rm AU \, (m_{\rm planet}/(3 m_{\rm star}))^{1/3}$, the depth of the smoothed potential is $r_{\rm s} = 0.1 r_{\rm H}$, outside of which the potential is the planets real gravitational potential. The simulation domain extends well beyond $5r_{\rm H}$ into the circumstellar disc. Gas is accreted from the circumstellar disc and onto the planet and the circumplanetary disc via the action of gravity, angular momentum conservation and cooling. The dynamics in the disc is solved via the Fargo-algorithm \citep{masset2000} in the code \textsc{FargOCA} \citep{bitsch2014, lega2014}. We employ viscous heating throughout the simulation domain with a constant physical viscosity, proportional to a viscous $\alpha$ of $\alpha \approx 3\times 10^{-3}-10^{-2}$.
The cooling is computed via radiation transport in the flux-limited diffusion approximation \citep{levermore1981} with the flux-limiter by \cite{kley1989}, for which $\kappa_{\rm R}$, together with the gas density $\rho$ and a fixed dust-to-gas mass ratio of $\epsilon=0.01$, plays the role of determining the optical depth $\tau$ of the dust and gas mix. Our previous analysis \citep{schulik2019} shows that a minimum resolution requirement must be observed in order for all those effects to play together and let accretion continue in a physically self-consistent manner, without numerical overheating of the envelope. }

Those simulations show that a low  $\kappa_{\rm R}$
is crucial for the formation of structures in protoplanetary gas envelopes  {and for the gas to reach Keplerian rotation}. This can lead to a flattening of the gas into circumplanetary discs and rotation velocities of the gas surpassing $v_{\theta}/v_{\rm Kepler}> 90\%$ for values of $\kappa_{\rm R}=0.01\rm\, cm^2/g$, while for high opacities of $\kappa_{\rm R}=1.0\rm\,cm^2/g$ the gaseous envelope remains fairly spherical and unable to form a rotating disc  {consistent with \cite{judith2016}}. We use the density-temperature data from the runs with those two constant Rosseland opacities for our synthetic observations.
Nominally, the high $\kappa_{\rm R}$ simulation will be matched with the 'S' dust mix (sim 1) and the low $\kappa_{\rm R}$ will be matched with the 'L' dust mix (sim 2).

\begin{figure}

	\centering
	\includegraphics[width=0.50\textwidth]{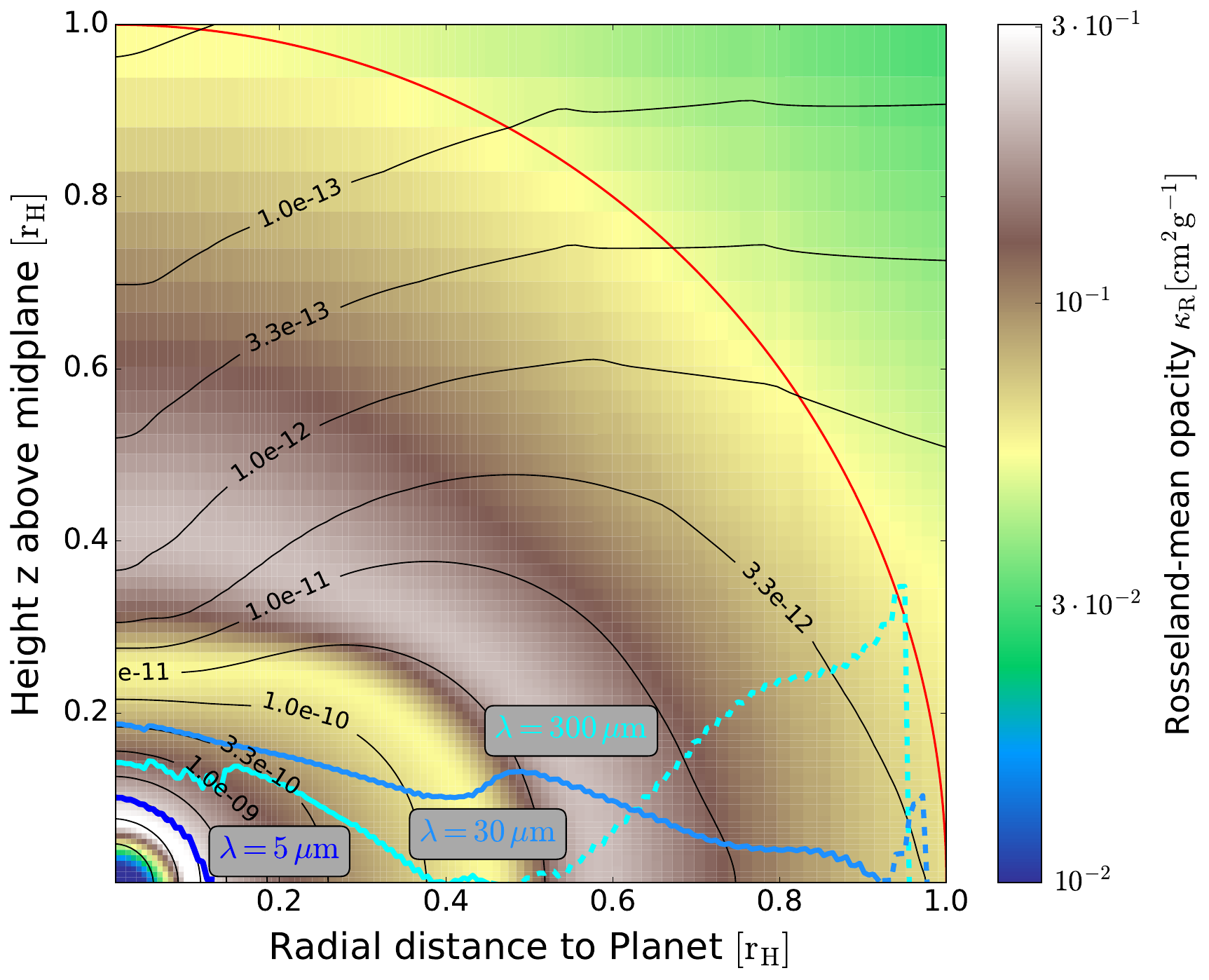}
   
\vspace*{+0.20cm}
\caption{Cylindrically averaged radial-vertical map of $\kappa_R$ computed only for ice and olivine components, centered on the planet for sim 1. This highlights the approximate 3-D structure of ice and olivine sublimation ellipsoids in the envelope, as seen via the corresponding opacity drops. Overlaying black contours denote the gas density in $\rm g\,cm^{-3}$ to highlight the disc-structure of the hydrodynamic solutions. The Hill-sphere is marked with a red circle. The actual $\tau=1$ surfaces at $\lambda = (5,\,30\,, 300)\, \mu$m computed with the full solution for sim 1 and viewing from $z=+\infty$ are overlaid in (darkblue, lightblue, cyan). Dashed colours denote the mirrored continuation of the surfaces below the disc midplane. The $\tau=1$ surfaces help to interpret spectral information.}
	\label{fig:opacity_map}
\end{figure}

\begin{figure}

	\centering
	\includegraphics[width=0.5\textwidth]{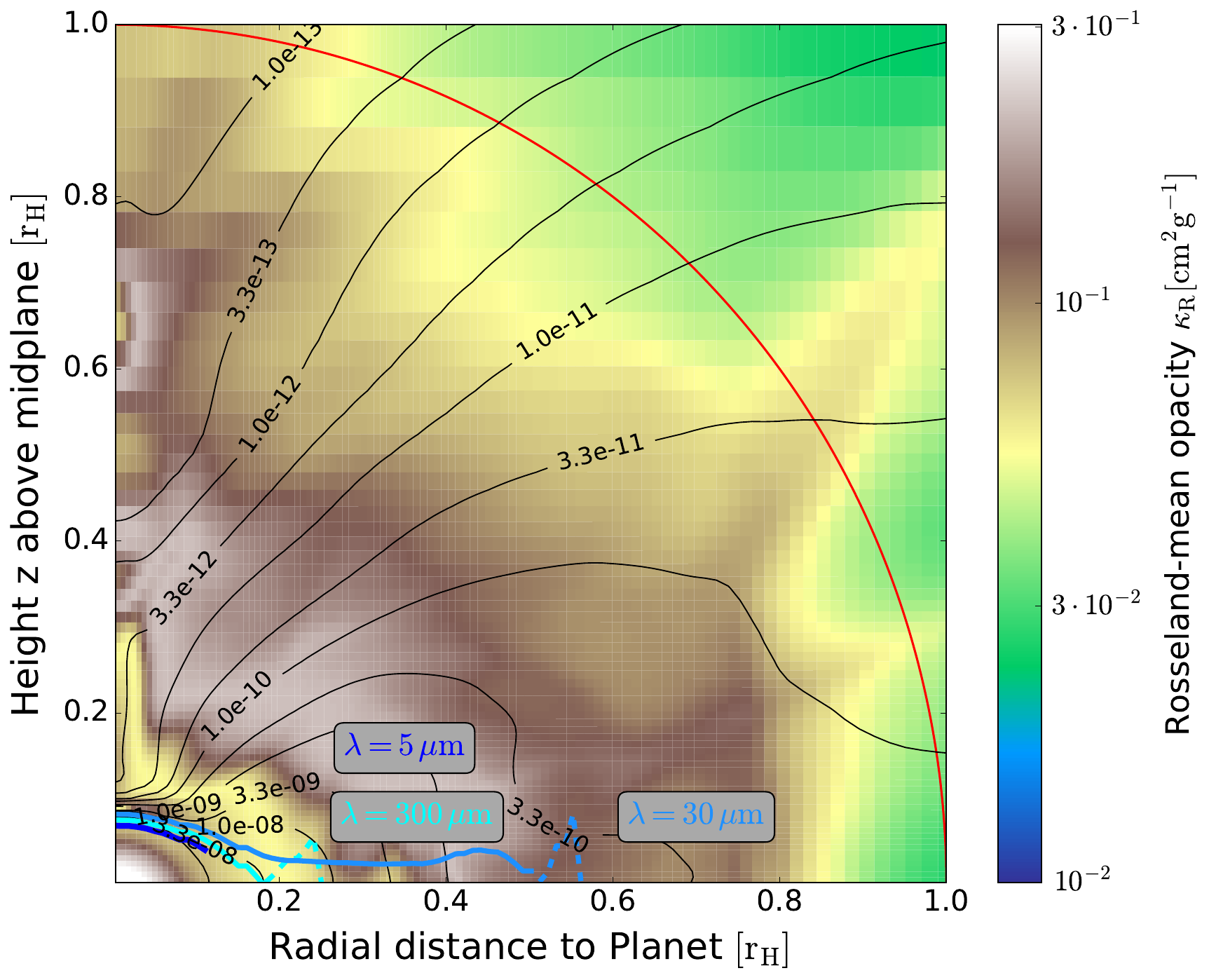}

\vspace*{+0.20cm}
\caption{Same as, Fig. \ref{fig:opacity_map} but for the low opacity hydro simulations and sim 6,  {with} having the same optical depth as sim 1 through a $\times$100 density increase. Again, the background colour is the Rosseland opacity computed only with Olivine and water ice to highlight compositional differences. The $\tau=1$ emission surfaces are again computed with the full opacity model including all species, as used in the paper to compute spectra. }
	\label{fig:opacitymap_sim6}
\end{figure}

 {
The structure of $\kappa_{\rm R}$ in our post-processed radmc3d runs is shown in Fig. \ref{fig:opacity_map} and reflects the temperature structure of the accreting const-$\kappa_{\rm R}$ gas giant envelope: Vertically above the planet, the accreting gas shocks and increases the temperature locally to $\sim$$500$K, leading to a deviation of the opacity and density contours. This effect is more extreme in the strongly cooling simulation with the low average $\kappa_{\rm R}=0.01\; \rm cm^2/g$, which can be seen in Fig.~\ref{fig:opacitymap_sim6}. There, a hot funnel of gas empties out the icy dust above the accretion shock, making it transparent to escaping radiation. The relative patchiness of the opacity structures seen in Fig.~\ref{fig:opacitymap_sim6} results from temperature fluctuations due to many complex adiabatic compression features and colliding flows in that simulation. In the midplane for both simulations, the density and opacity contours are aligned, as no disequilibrium radiation from the shock is present in that region. We note that in \citep{schulik2020} we also ran a simulation with non-constant opacities given by \cite{belllin1994}, which would reproduce this radial-vertical structure of $\kappa_{\rm R}$ more closely. However numerical speed and stability issues favoured our usage of the const-$\kappa_{\rm R}$ hydro data, also noting that the gas dynamics in the outer envelope does not change significantly. }

In order to quantify the influence of the CPD on the overall spectra,  {i.e. to assign 'disc-dominated' or 'planet-dominated' wavelengths,} we first generate a spectrum of the pair CPD + planet and then substract the CPD region from the data in a second step. Both those steps are performed in a face-on view (from the top of the disc) because otherwise one would peer sideways into the planet after subtracting the CPD, which would add excess contributions to the spectrum. This approach is in contrast to the synthetic image in Fig. \ref{fig:observation_image}, which is taken at a $50^{\circ}$ inclination to mimic the view on PDS 70c.

The net effect of lowering $\kappa_{\rm R}$ on the optical depth via dust growth can be achieved by either increasing the dust sizes or by decreasing the overall number of dust grains. Hence we also process the $\kappa_{\rm R}$ hydrodynamic data with the 'S' and 'ISM' dust grains, but with 100 times (1000 times) lowered dust density, giving rise to sim 3 (sim 4). We call sims 1-4 the 'early' type sims due to their low overall masses.
As outlined in \cite{schulik2020} our simulations feature only the initial few orbits of the formation stage of a CPD. Therefore our CPD dust masses are very low (on the order of a few $10^{-5}\,m_{\oplus}$). In order to obtain spectra of 100 times more massive CPDs, like those estimated in \cite{isella2019}, we simply scale all gas densities from sim 2 up by a factor of 100, and again use the different possibilities for obtaining a low $\kappa_{\rm R}$, giving rise to sims 5-7, which we call 'late'-type sims. 

Additionally, under the hypothesis that 'late'-type CPDs might be significantly cooled down compared to our simulations, we investigate another set of 'late \& cool' CPDs, which feature a factor of 3 reduction in temperature for all cells. This procedure gives rise to sims 8 \& 9. The late and cool sims are only processed with 'L' and 'ISM' dust mixes.

In order to simulate the effects of sublimation, we assume that $\rho_{i,D}=0$ if the local temperature in a grid cell is above the sublimation temperature for that dust species $ T_{i,\rm sub}$. Ice coating sublimation is approximated by setting each dust species to their nominal values $\rho_{\rm Ol}$, $\rho_{\rm C,major}$, $\rho_{\rm C,minor}$ as described above, and $\rho_{\rm H_2O,D}=0$ at dust temperatures $T>T_{\rm H_2O,\rm sub}$. The temperature for all dust species is taken as equal to the gas temperature.

\begin{figure}

	\centering
	\includegraphics[width=0.50\textwidth]{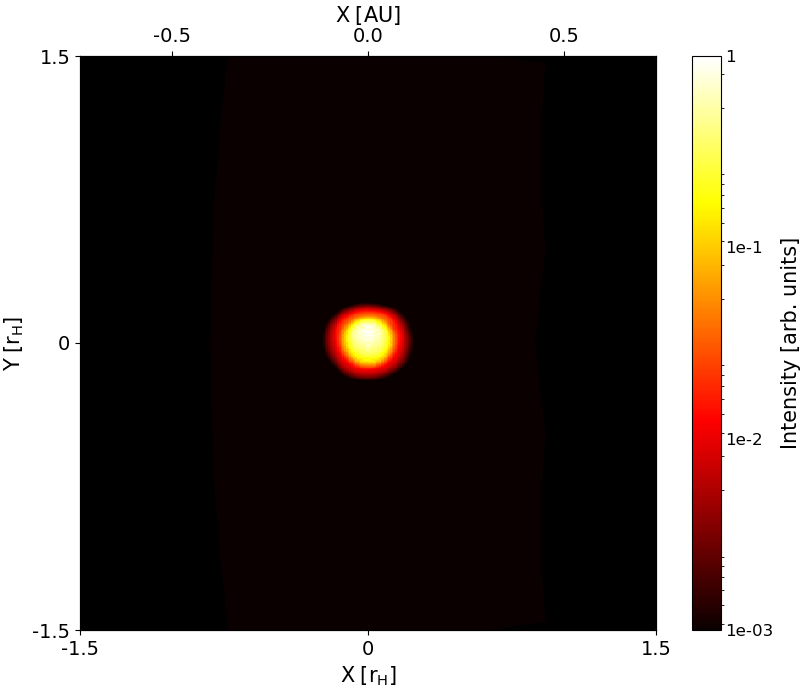}
   
\vspace*{+0.20cm}
\caption{Intensity map of a synthetic observation of the planet and CPD at 5 $\mu$m. The image is as expected from the computed 5 $\mu$m-emission surface in Fig. \ref{fig:opacity_map}. The CPD is not visible, as at those wavelengths water ice and organics as opacity sources have sublimated away. This image is also representative of any simulation setup where we find planet-dominated spectral regions, which can also happen up to the sub-mm. It is then usually the blackbody-tail from the planetary contribution overshining the CPD signal. The latter is then weak, as the flux is $F_{\rm \lambda}\propto \kappa(\lambda) \Sigma$, with $\Sigma$ being the surface density, and $\kappa(\lambda)\propto \lambda^{-2}$ in the far-infrared. 
Increasing the contrast to arbitrary, unrealistic levels would still make the CPD contribution visible (see text for more details).}
	\label{fig:observation2}
\end{figure}

\section{Synthetic observations}
\label{sec:observations}


\subsection{General considerations - Emission surfaces leading to spectra}

We begin our analysis with an investigation of the emission surfaces for simulation 1. The emission surfaces are computed by integrating the optical depth from infinity in a top-down view onto the CPD per individual wavelength. The surfaces are then found where the optical depth becomes larger than unity. They are a useful tool to understand the contribution of the CPD to the total spectrum, as well as the overall shape of the spectrum. This is due to the fact that an existing emission surface will pose a very bright area, radiating as blackbody, whereas regions without emission surface, i.e. optically thin regions are not very bright \cite{mihalasmihalas}.

In Fig. \ref{fig:opacity_map} we show a background colour map of $\kappa_{\rm R}$, computed with only the olivine and water ice components for simulation 1, in order to clearly show the compositional transition at the water and olivine sublimation lines in the envelope of that particular giant planet. It becomes evident that the water ice line is in general more shaped like an ice ellipsoid. We overlay the gas density contours in black, which illustrate that the hydrodynamic data for $\kappa_{\rm R}=\rm 1\; cm^2/g$ has difficulty forming a CPD. The hydrodynamic data used for sims 2-9 forms a flat CPD (which we show in Fig. \ref{fig:opacitymap_sim6} for completeness), but we found sim 1 to be an ideal illustrative example to study the emission properties of the giant planet envelopes.

Next, we show the emission surfaces at $5,30$ and $300 \, \mu$m now computed with the full opacity model, i.e. all dust species included, in Fig. \ref{fig:opacity_map}. It becomes evident that the emission surfaces (particularly around 30 $\mu\rm m$) can follow a disc-like structure, owing to a significant opacity contribution from water ice and carbonates. The emission surfaces are representative of the final synthetic image at a given wavelength.

The blue line, representing the emission surface at 5 $\mu$m in Fig. \ref{fig:opacity_map}, will be planet-dominated, as for temperatures corresponding to this wavelength ($T\approx$ 600 K $> T_{\rm sub,C}, \; T_{\rm sub,H_2O}$) both organics and water have sublimated. This results in a rather unspectacular appearance in the 5 $\mu$m image, Fig. \ref{fig:observation2}, and leaves the silicate disc outside of $r>0.1\,r_{\rm H}$ transparent, as the silicate has an opacity minimum at this wavelength, as previously seen in Fig. \ref{fig:opacities} (\textit{Left}). 

\begin{figure*}	
\hspace*{+0.20cm}
  \begin{subfigure}{0.50\textwidth} 
	\centering
	\includegraphics[width=\textwidth]{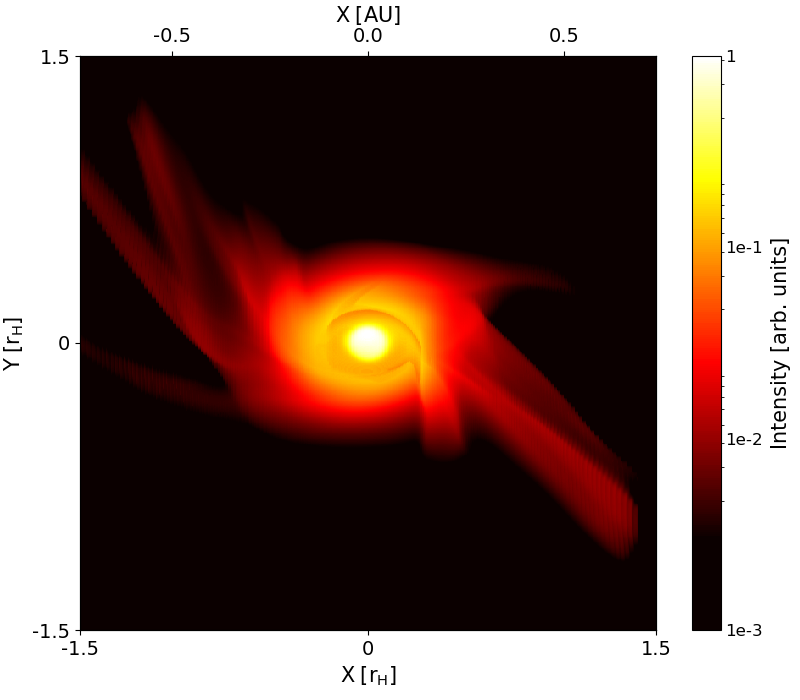}
   \end{subfigure}
	\begin{subfigure}{0.50\textwidth} 
	\centering
	\includegraphics[width=\textwidth]{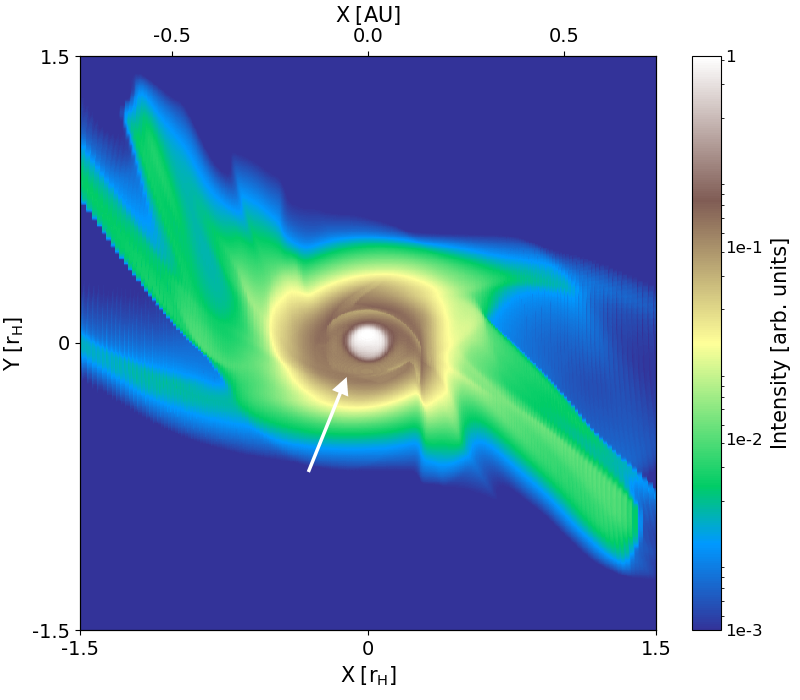}  
   \end{subfigure}%

\vspace*{+0.20cm}
\caption{Synthetic intensity maps for the planet and CPD for sim 1 at 30 $\mu$m seen at 50$^{\circ}$ inclination, to mimic PDS 70c. Those maps assume a spatial resolution far superior to ALMA, but serve the purpose to illustrate the intensity contributions to the spectral flux at this wavelength. On the left we show a intuitive heat map, on the right with a more contrasted colour map. On the right, one can distinguish several regions according to their emission intensity $I$: Emission from the innermost planet, originating from olivine, $I\approx 1$, the CPD with the water iceline $I\approx0.1-0.6$, also marked with the white arrows, the outer edge of the CPD $I\approx 0.03$, various disc structures including spiral arms at $I\approx 0.01$. The arrow marks the water ice sublimation line, which through its associated drop in opacity reveals some inner parts of the envelope. }
	\label{fig:observation_image}
\end{figure*}

The light blue line, showing the emission surface at 30 $\mu$m, results as a more complex combination of all opacity carriers involved. As this emission surface clearly cuts through the water ice sublimation ellipsoid, a synthetic image at this wavelength merits some further study. In Fig. \ref{fig:observation_image} we show such a synthetic image.

\begin{figure*}	
\hspace*{+0.20cm}

\includegraphics[width=0.29\textwidth]{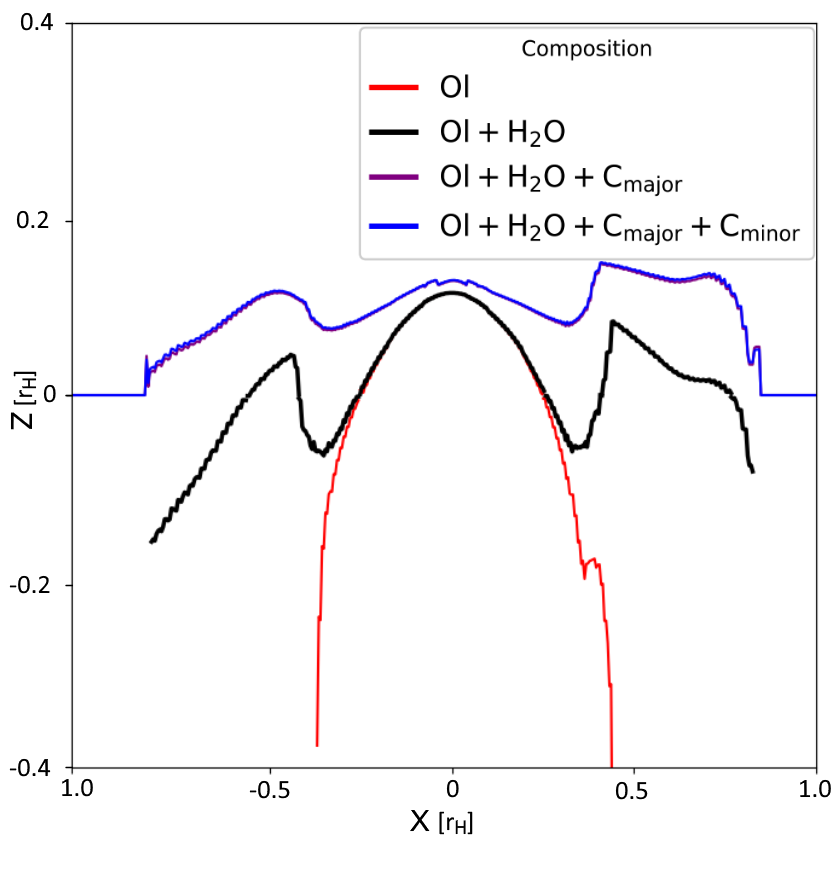}
\includegraphics[width=0.34\textwidth]{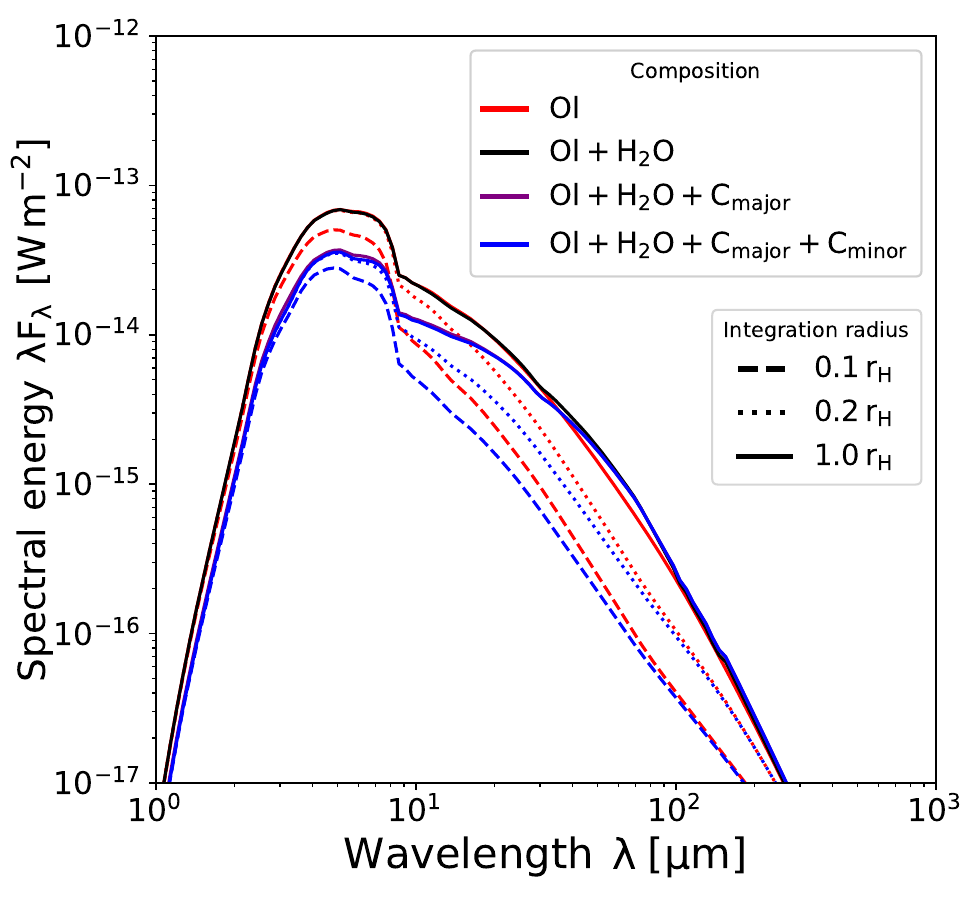}
\includegraphics[width=0.34\textwidth]{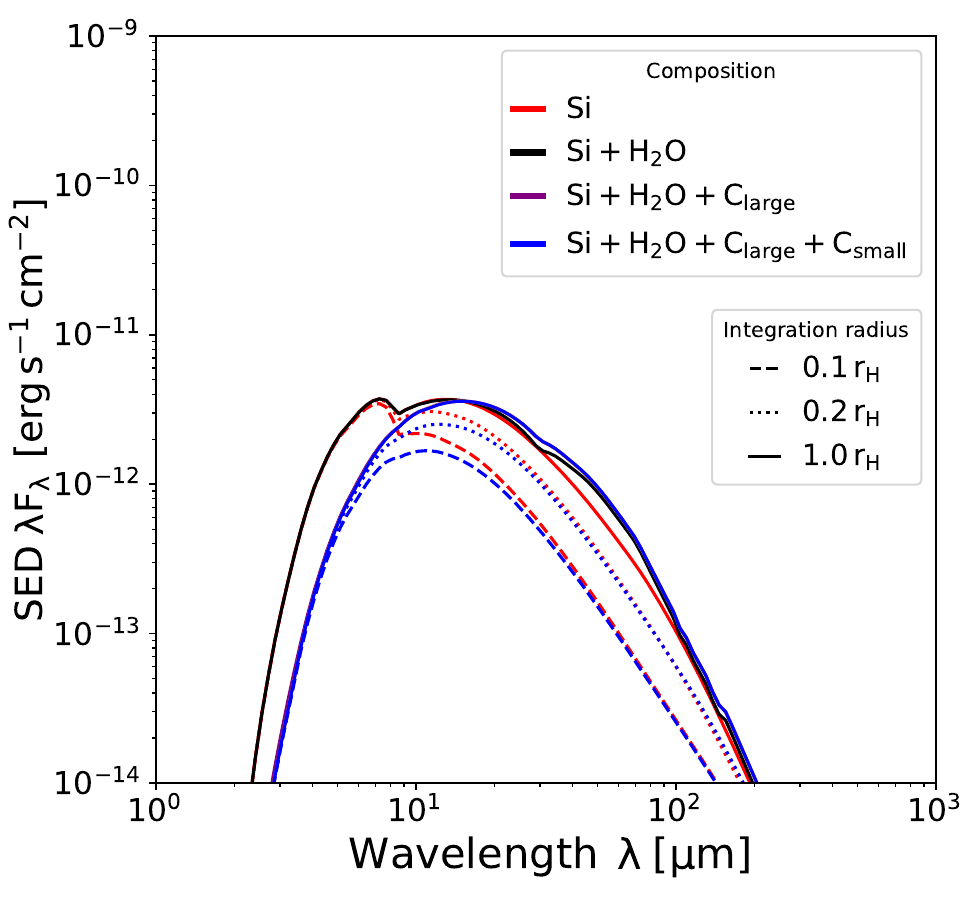}

\vspace*{+0.20cm}
\caption{ {Left and middle:} Effects of successively adding opacity carriers into sim 1. We show the emission surfaces in real space at 30 $\mu$m on the left, again integrated from $z=+\infty$, with the radial distance on the $x$-axis and the vertical distance on the $z$-axis. The resulting full spectrum is shown on the right.
The Olivine (Ol) sets the dominant part of the spectrum, with the contribution from the water ice being minuscule mainly due to being outshone by the Olivine close to the planet. Once the organics components are added ($\rm C_{major}$ and $\rm C_{minor}$), the central emission surface is lifted up to colder temperatures (longer wavelengths), leading to a weaker signal in the NIR through extinction. Interestingly, the carbon greys out its own sublimation signature in the emission surface, whereas the sublimation line is clearly visible for water in the emission surface, but only weakly in the spectrum. 
When measuring the CPD contribution to the spectrum, we subtract everything outside of $0.1 r_{\rm H}$ from the simulation domain. This process can be seen on the right side, when following the blue dashed line as it changes into the dotted blue and then the solid blue line. It becomes evident that this spectrum, when all dust ingredients are included, gains a lot of flux for $\lambda>10 \,\mu$m, while its gain is minor for $\lambda<10 \,\mu$m. This leads to the conclusion that for $\lambda>10 \,\mu$m this model is CPD dominated. The same can be repeated for the Olivine-only spectra. Those inform us that already the hot olivine blob seen on the left can be counted as CPD contribution.   {Right:} Same as on the left side, but for sim 6, deriving from the low $\kappa_{R}$-data, but having the same optical depth as sim 1 through a $\times$100 density increase. Water emission from the CPD plays a larger role in this spectrum, particularly between 30 $\mu$m and 100 $\mu$m. 
}
	\label{fig:surfaces_chemistry_merged}
\end{figure*}

This image is generated at an angle of $50^{\circ}$ similar to PDS 70c, in two different colour maps, to discuss the contrast of features seen in them. A typical 'heat' colour map, shown in Fig. \ref{fig:observation_image} (\textit{Left}), carries intuitive information pertaining to the relative brightness of features, but does not clearly show the CPD without hiding surrounding features. Therefore we also show in Fig. \ref{fig:observation_image} (\textit{Right}) an adjusted colour map, which can more clearly guide the expectations for the CPD contribution to the 30 $\mu$m spectrum. The CPD shows at its inner edge the water sublimation front, decreasing its brightness (marked as white arrow). However, radially outwards from the planet, behind the water iceline, the CPD has a brightness of about 30\% of the maximum value and a much larger surface area. Hence, for this wavelength we would expect the spectrum to have significant CPD contributions. This is not always given by the water ice, as also olivine from inside the water iceline can contribute importantly to the CPD contribution. Additionally to the planet and the CPD, other dense flow features radiate at this wavelength, such as the circumstellar spiral arms. Those spiral arms additionally show their complex 3-D structure due to the inclination of the image.

The cyan line at 300 $\mu$m interestingly leaves the CPD midplane mostly transparent. This is because sim 1 uses the 'S' particles, which exhibit the opacity slope in this wavelength region, as seen in Fig. \ref{fig:opacities} (\textit{Left}). Only a weakly radiating CPD component from an emission surface below the midplane contributes to the spectral intensity at this wavelength.
This is an important point of our analysis, as it explains why the CPD emissions can be dominated by the planet in the sub-mm for small particles, but the CPD contribution can be very substantial when using large particles at those long wavelengths.

As the 30 $\mu$m wavelength image turns out to exhibit significant CPD features, we now first continue analysing its compositional build-up, before turning to a joint analysis of composition and spectra.

\subsection{The full model - Contributions of individual dust components to the emission surfaces and the spectrum}

We next investigate the relative importance of the individual dust species in forming the emission surfaces and spectra. To do this, in Fig. \ref{fig:surfaces_chemistry_merged} (\textit{Left}) we show exemplary emission surfaces at 30 $\mu$m, and how they successively build up as we add the dust components, for the 'S' mixture in sim 1, which has maximum dust sizes of 10 $\mu$m. This is complemented by the spectrum of the entire object integrated over varying fractions of the Hill-radius, in Fig. \ref{fig:surfaces_chemistry_merged} (\textit{Middle}), which inform about the relative importance of the surface area w.r.t the spectrum.

Olivines, sublimating at the highest temperatures, form the dominant opacity source near the planet. The olivine spectrum exhibits a peak shortwards of 10 $\mu$m, because the olivine opacity there decreases (see Fig. \ref{fig:opacities}) and hence opens a transmission window into higher temperature regions of the protoplanetary envelope.
Note also how the olivine contribution remains important in this spectrum when including more and more surface area. This can be seen in Fig. \ref{fig:surfaces_chemistry_merged} (\textit{Middle}), when going from the dotted red, to the dashed red, to the solid red line, it becomes evident that sim 1 exhibits an olivine dominated spectrum.

After adding water, the emission surface profile is only slightly changed inside the water-ice ellipsoid, due to contributions from water above the planet. Outside the midplane water-ice ellipsoid, there are strong contributions evident from the water ice, which then extends the emission surface nearly out into the circumstellar disc. In the spectrum, as water can exist only at cold temperatures, there is a 20\% increase in emissivity at wavelengths longwards of 30 $\mu$m, but no strong water features are visible. This is due to the enormous surface area of the central olivine blob, which is outshining all the other contributions. For the same analysis of our low $\kappa_{\rm R}$-hydro data, but with the same particle sizes and optical depth, see  { Fig. \ref{fig:surfaces_chemistry_merged} (\textit{right})}.

With the carbon added, there is a fairly uniform increase in optical depth, due to the grayness of the carbon spectra. Interestingly, the carbon sublimation at 410 K leaves only a very minuscule imprint onto the emission surface, which may be surprising, and the spectrum overall. The main effect of the carbon on the spectrum is to move it to slightly colder temperatures and longer wavelengths, and provide extinction for the underlying olivine bump in the spectrum.

The general build-up of any spectrum will follow the same principles as discussed here. Hence, for a low CPD contribution in any part of the spectrum, one can be sure that the emission surface will be similar to the red curve in Fig. \ref{fig:surfaces_chemistry_merged} (\textit{Left}) and vice versa.

We now turn to compare the 'S' and the 'L' dust mixture, with their corresponding hydrodynamic models.

\subsection{Early type Spectra - sim 1 and sim 2 and CPD contributions}

We now present an analysis of fractional flux emitted from the planet compared with that from the CPD for sims 1 and 2. Those two simulations, as discussed earlier, were generated with average $\kappa_{\rm R}$ differing by a factor of 100, hence their average optical depths will vary accordingly, and so will their relative CPD contributions to the spectrum.

As can be expected from the preceding discussion, the CPD emission will contribute significantly to the total emission at wavelengths at which the CPD region is optically thick. Hence, when the emission surfaces cut through dense regions, including the hot midplane, we find CPD dominated spectra. 
In Fig. \ref{fig:spectra_lowvshigh_merged} \textit{(Left)} we see 
 {how the CPD contribution dominates the blue curve due to the the high $\kappa_{\rm R}$ contribution from small particles. }
A grey envelope around the  {plotted} spectrum indicates the CPD flux contributing 10\% or more of the total spectrum, and a thick coloured envelope around the spectrum indicates that the CPD is a dominant contributor  {($>$$50\% $)} to the spectrum.
It becomes evident that the CPD is generally dominating the spectrum longwards of 10 $\mu$m until $\sim$300$\mu $m, as should be expected by our preceeding discussion. Longwards of $300\;\mu$m, the CPD becomes transparent due to the rapidly falling Mie-slopes in the opacity functions of small particles, and the corresponding disappearance of emission surfaces. The spectrum shortwards of 10 $\mu$m is planet-dominated, due to the minimum in olivine opacity, and the fact that the opacity carriers in the CPD evaporate away at higher temperatures.

Simulation 2, having a low average $\kappa_{\rm R}$ and the corresponding 'L' dust population with a maximum size of 1 mm, is shown as the solid red curve.
As seen previously, in Fig. \ref{fig:opacities}, particles of this size are generally very grey, and different dust species only differ in their constant opacity level, with water ice having the strongest contribution, then the carbonates, followed by olivine. However, due to the overall lower level of opacity, the CPD is transparent at all wavelengths, as can be seen in Fig. \ref{fig:spectra_lowvshigh_merged} \textit{(Left, red curve)}. 

There is still a weak ($<10\%$) contribution to the overall width of this spectrum from some radiation in the CPD, as can be seen in Fig. \ref{fig:spectra_lowvshigh_merged} \textit{(Left)}, through comparing the red curve with the width of black-body (3). Hence, it can be dangerous to assign the total spectral flux to a CPD, only because the CPD is, or is believed to be, transparent.

\subsection{PDS 70c and black-body fits}

\begin{figure*}	

	\centering
	\includegraphics[width=0.45\textwidth]{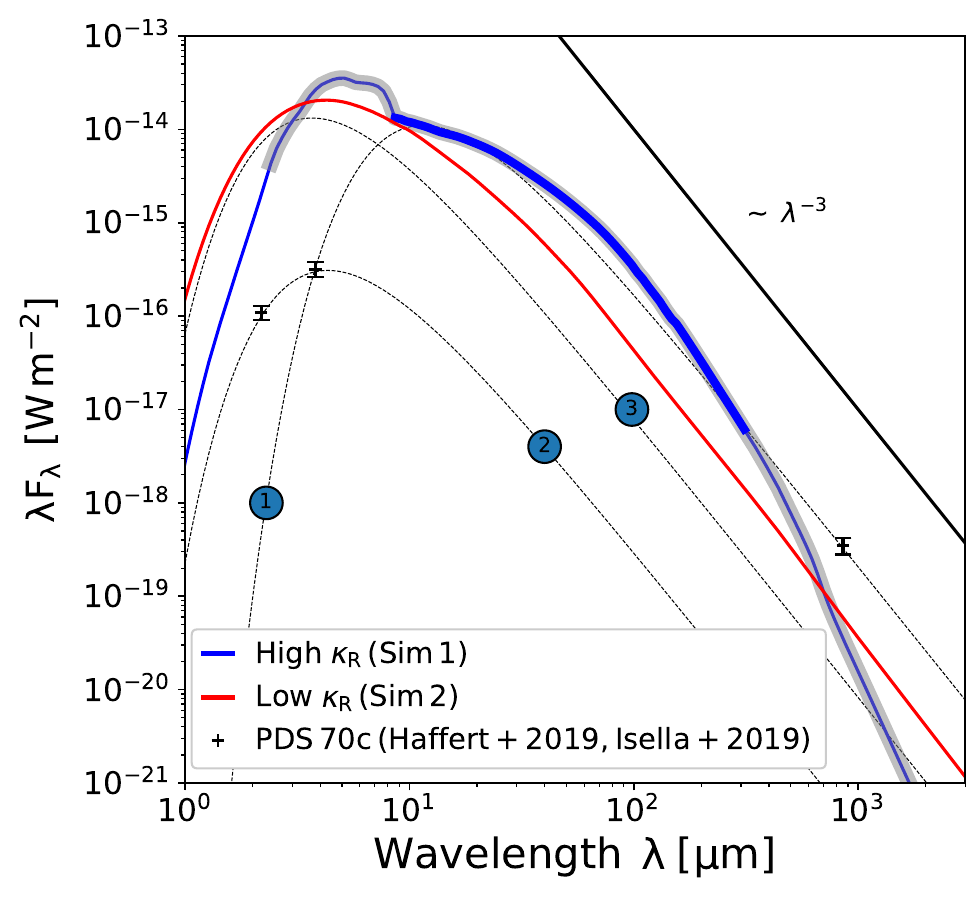}
    \includegraphics[width=0.45\textwidth]{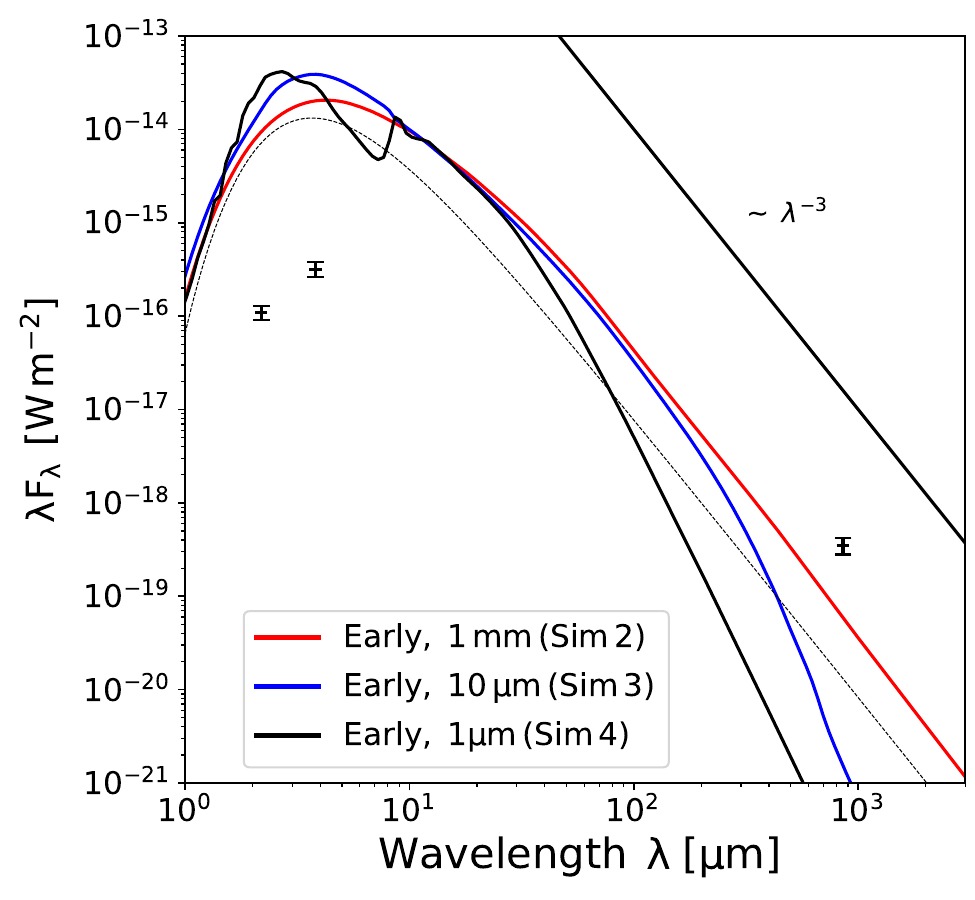}
   
\vspace*{+0.20cm}
\caption{ {Left}: Spectral energy distributions for sims 1 and 2 with all dust species included and normalized to the distance of PDS 70, 113 pc. Furthermore, we plot the fluxes for PDS 70c in the Ks (2.18 $\mu$m), L' (3.8 $\mu$m) and 855 $\mu$m bands reported by \cite{haffert2019, isella2019}. For conversion details and flux errors see Appendix A.
The observational data as well as our model spectra are compared in width to black-bodies (BB), plotted as thin, dashed curves. There are BB (1) ($T=$300K), BB (2) ($T=$850K) and BB (3) ($T=$1000K). PDS 70c fails to be modeled by BBs (1) or (2) and due to this width, it reveals to possess a CPD component. Our model data compared with BB (3) shows it to have multiple BB contributions. Grey shading indicates a weak CPD contribution, i.e. $F_{\rm CPD} > 0.1 F_{\rm Planet}$ and thick lines indicate a dominant CPD contribution, i.e. $F_{\rm CPD} > F_{\rm Planet}$.  {Right}: Spectra for the low $\kappa_{R}$-simulations, emphasizing the effect of dust growth on the spectra. In order to keep the optical depth constant for the smaller dust particles with higher opacities, we adjusted $f_{\rm DG}$ downwards accordingly. For all wavelengths, the CPD is transparent and the emission surfaces are convex features being situated around the planet, similar to the olivine-only case in Fig. \ref{fig:surfaces_chemistry_merged}\textit{(Left)}. The comparison of spectral width to the black-body (dotted black line) shows that for the mm-particle case, the spectrum is a convolution of black-bodies over the 3-D structure, while the origin of the width is less clear for the smaller particles. Furthermore, the growth from ISM-like towards mm-sized particles 1) greys the spectrum and 2) produces a FIR-tail that is significantly shallower than the expectation from single-temperature, single-small-particles, which would be a $\sim\lambda^{-5}$-slope.  }
	\label{fig:spectra_lowvshigh_merged}
\end{figure*}

In Fig. \ref{fig:spectra_lowvshigh_merged} we also show the data available at the moment from broad-band photometry and sub-mm imaging for the planet and the putative CPD around PDS 70c. For details on the conversion process  {of data}, see Appendix A. The wide spacing of the PDS 70c data seems to be already indicative of an at least two-black-body solution, when comparing black-body 1, which connects the L'-band and sub-mm point of $T=$300 K, and black-body 2, connecting the Ks and L' band points with $T=850$ K. It seems, however, implausible that a CPD would be fitted by only one black-body. 

Additionally, the planetary black-body fit can inform us about the true planetary radius (not the emission-surface radius in our hydro models), by converting the integral of the blackbody flux into a surface area, which would correspond to $R_{\rm P}\approx (4.5 \pm 0.5) R_{\rm Jup}$. This radius is too large to be explained by current formation models (\cite{mordasini2017}) at the age of PDS 70 for any planetary mass, {and at odds with the findings in \citet{wang2020}. Further, \cite{christiaens2024} caution that the nearby bands (Y, J, H) to the ones used in this work (Ks, L') suffer from contamination due to scattering stellar light. A hotter planetary temperature, an even stronger dusty contribution to the spectrum than inferred here or stellar scattering are all possibilities to solve the discrepancy in planetary radius.}

Hence, we will next investigate how particle growth influences our model CPD spectra, with a focus on the spectral slope in the far-infrared. Once this is understood, we discuss whether it is possible to improve the planetary Ks and L'-band black-body fit by substracting a reasonable CPD model.

\subsection{Early type spectra - varying dust sizes, while keeping $\tau$ constant for sims 2-4}

We focus now on the low $\kappa_{\rm R}$ simulation in order to eliminate complications due to changing density-temperature structures, when changing $\kappa_{\rm R}$. We stress the possibility to generate the same optical depth $\tau$ at low $\kappa_{R}$ with particles of smaller sizes, despite their higher opacities. For this, we decrease the dust-to-gas ratio accordingly, in order to leave the approximate optical depth overall the same. Specifically this means that when using the 'S' mixture with a maximum size of 10 $\mu$m, which has a $\sim$100 higher opacity, the dust-to-gas ratio was adjusted to $10^{-4}$, and accordingly even lower for the ISM dust.

In Fig. \ref{fig:spectra_lowvshigh_merged} \textit{(Right)} the changes in the spectrum when growing the dust particles becomes evident. Due to the low density and low average opacity the CPD remains transparent for all dust sizes, and the emission becomes planet-dominated. The 'ISM' particles of 1 $\mu$m size show essentially a pure olivine spectrum in emission, but its contrast is greyed out as for sim 1 through carbon. In the olivine-opacity minimum between of 3-10 $\mu$m (as seen in Fig. \ref{fig:opacities}), the entire envelope is transparent, as the opacity there is low enough to reach the olivine sublimation line.
The slope of the spectrum in the FIR/submm region is in general the sum of all radiating black-body contributions $B_{\lambda}$ weighted with their opacities, $\lambda F_{\lambda} \propto \sum \kappa_{\lambda} \lambda B_{\lambda}(T)$ (e.g. \cite{backman1992}). A single, optically thick black body would have $\lambda B_{\lambda} \sim \lambda^{-3}$ and with the opacity slope for small particles being $\kappa_{\lambda}\sim \lambda^{-2}$ in that region, it should be $\lambda F_{\lambda} \propto \lambda^{-5}$. As the spectrum for the small particles here measures however $\lambda F_{\lambda} \propto \lambda^{-4.3\pm0.1}$, we can attribute this flatter slope to the summation effect of black bodies to the spectrum.

For the dust 10 of $\mu$m in size, the carbon plays an important effect in greying out the olivine minimum shortwards of 10 $\mu$m.
This renders the entire dust mix near the planet optically thick  and the olivine features appears only very weakly as bump in emission. The spectrum has changed its slope, in the FIR, with a break at $\sim$300 $\mu$m, corresponding to the change in opacity slope at $\sim$100 $\mu$m, as seen in Fig. \ref{fig:spectra_lowvshigh_merged}\textit{(Right)}.

Finally, the mm-sized particles show a featureless convolution of black-bodies, with slope slightly steeper than $\lambda^{-3}$, of about  $\lambda^{-3.2\pm0.1}$. This is because in the sub-mm the mm-particle opacities show a weak slope themselves.

\subsection{Late type spectra and a fit for PDS 70 c}

We now turn to investigate the spectra of 'late' type CPDs, for which we have used the low $\kappa_{R}$ hydrodynamic data, and multiplied the density by 100, in order to match the masses in our simulations with dust mass estimates for PDS70c from \cite{isella2019}.

The first three simulations use the temperature structure as given by the hydro data (sims 5-7), while we also artificially decrease the temperature in the entire domain by an arbitrary factor of 3, in order to imitate a massive, cooled down disc (sims 8, 9).

\begin{figure*}	
	\centering
	\includegraphics[width=0.45\textwidth]{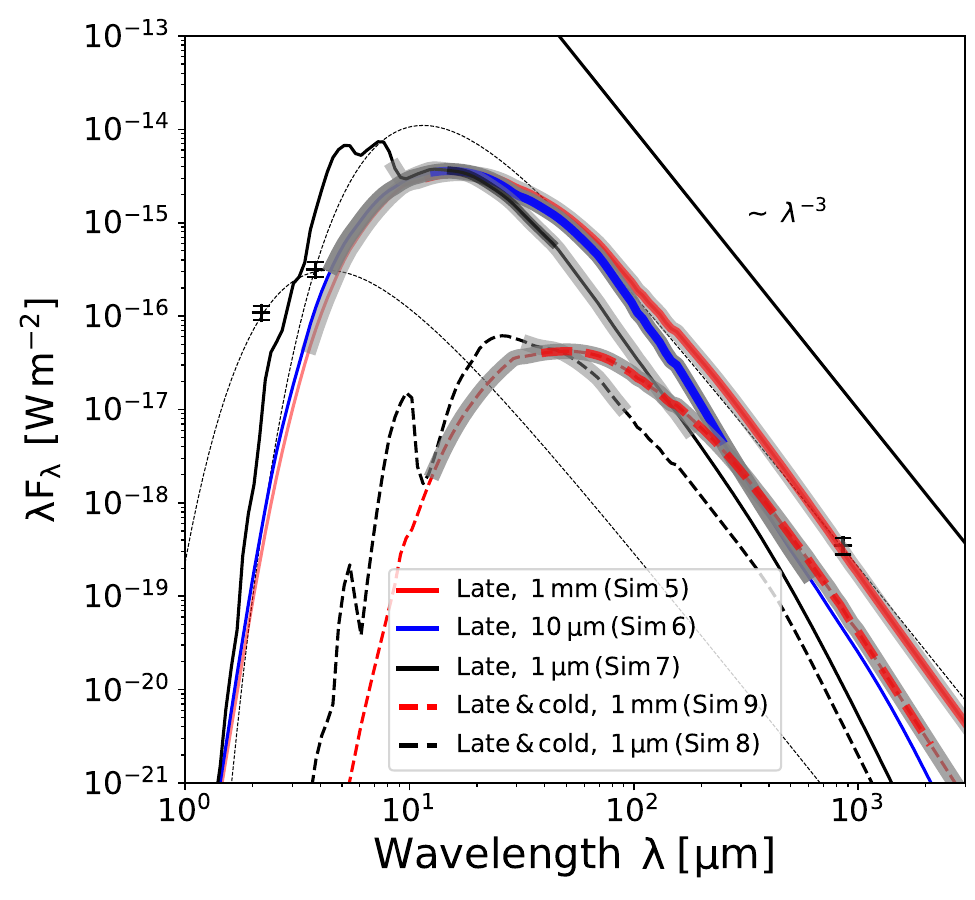}
 	\includegraphics[width=0.45\textwidth]{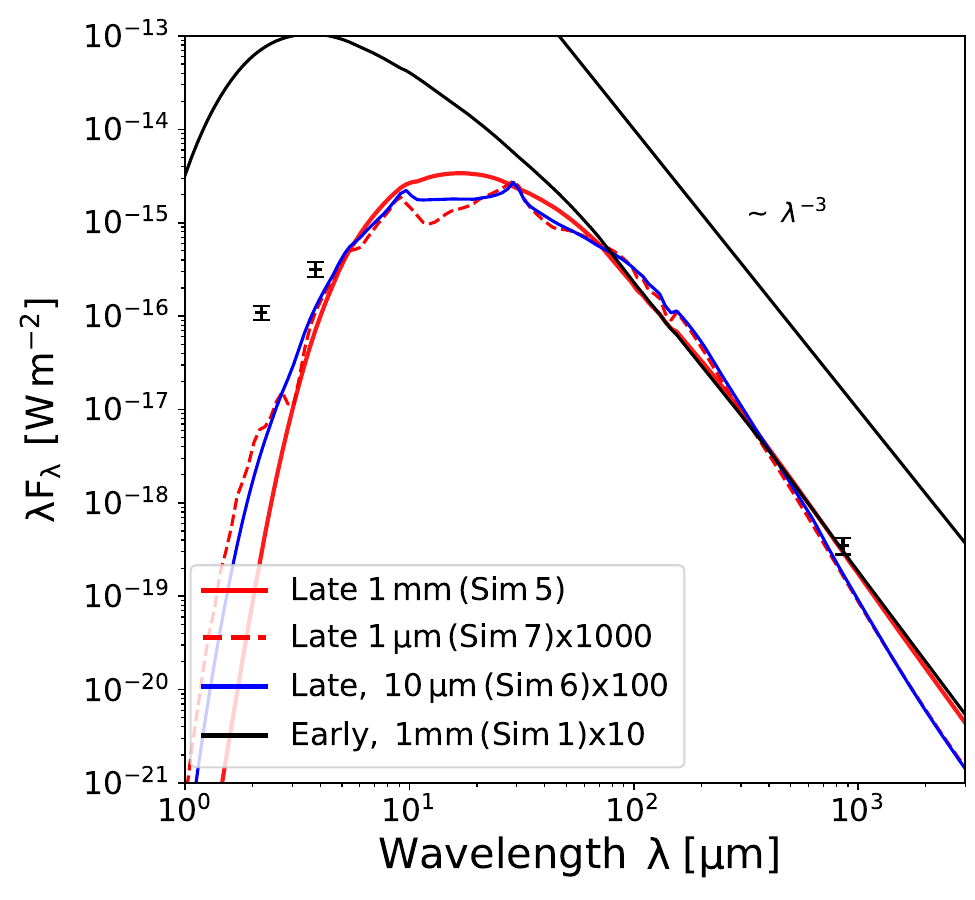}
\vspace*{+0.20cm}
\caption{  {Left:} The late models with $\times100$ times increased dust mass show significantly increased CPD fractions in their spectra (solid lines), due to the formation of emission surfaces in the CPD. We keep the previously introduced black-bodies for very rough expectations on the planet and CPD emissions.
FIR slopes are again black-body-like for mm-sized particles, while the overall steepness is not as dramatic as for the early-type small particles, making a fit more difficult. 
However, the width of the PDS 70c data can be approximately fitted when considering larger grains.
The artificially cooled down envelopes (dashed lines) are too cool to fit the observations and particularly undershoot the ALMA measurment.  {Right}: Degeneracy check for the fits on the left. We attempted verify whether increases in dust mass of smaller particles can also fit the seen width, which we can confirm here. Hence more data is needed to distinguish between mass and particle size. Curves are plotted without the indicators for the CPD fraction, as the viable fits all are optically thick in the CPD region and hence the sub-mm signal is actually emitted by the CPD.}
	\label{fig:spectra_late_merged}
\end{figure*}

From Fig. \ref{fig:spectra_late_merged}\textit{(Left)} one can read off that dust growth leads to an increasing CPD contribution to the total spectrum, mainly due to the formation of extended emission surfaces, as seen in sim 1, which can be attributed again to the greyness of the mm particles. This is an important finding, as it shows that also our simulation data for the low $\kappa_{R}$, which actually forms CPD structures with important keplerian rotation, can be detectable in the late stages of their evolution.

The FIR slope of the spectrum evolves weaker than for the early spectra, from around $\lambda^{-3.5\pm0.1}$ for the ISM particles to $\lambda^{-3.1\pm0.1}$ for the mm-particles. The less clear distinction between small and large particles stems from a stronger contribution of blackbody mixtures to the spectrum.

The 'cold' simulations naturally have less total emission, but the disc contributions remain dominant longward of 30 $\mu$m for the mm-sized particles (sim 9). From the preceeding discussion, it is not surprising that the ISM particles (sim 8) are essentially planet-dominated, but this is strengthened due to the density structure for the following reason. We seed opacity carriers in cold regions proportional to the gas density, this gas density is however severely depleted due to the ongoing accretion process. Therefore, although the number of ice particles is enhanced in the cold case overall, the optical depth does not increase significantly above the CPD due to ice carriers.

Finally, it seems that we can obtain a fit to the broadness of the PDS 70c data with a two-component solution, that of our CPD data for the mm-sized particles and that of a planetary black-body. From Fig. \ref{fig:spectra_late_merged}\textit{(Left)} it becomes clear that the spectra for the CPD + planet for the 10 $\mu$m (sim 6) and 1 mm sized particles (sim 5) are natural candidates to fit the width of the PDS 70c spectrum. We emphasize here that this is an immediate result of our mass increase that we performed in order to be consistent with the dust mass estimate for this source.

An inspection of the 10 $\mu$m  {dust-size} model spectrum in Fig. \ref{fig:spectra_late_merged} \textit{(Left, solid blue line)} reveals that this spectrum is planet-dominated at 855 $\mu$m and is therefore inconsistent with another planetary black-body, which would be needed to fit the Ks and L' datapoints. Only the mm-sized particles (sim 5) possess the required sub-mm component to fit the spacing between the NIR and sub-mm data, without the need for another planetary black-body.

If we take the mm-sized late model, subtract our model planet and instead insert a more realistic planetary black-body into it that would fit the Ks and Lp datapoints we would however still remain with an unrealistically large planet. A smaller, more realistic planet would require a stronger CPD contribution in the L' band, which we found challenging to model. A natural way to increase the CPD contribution in the L band would be an improved version of our hydrodynamic data, because, as we outlined in \cite{schulik2020}, our modeling overestimates the planetary contribution.

\subsection{Uniqueness of the fit}

In order to determine the robustness of the mm-sized particle fit, we attempted a number of other solutions to fit the width of the PDS 70c spectrum, as shown in Fig. \ref{fig:spectra_late_merged}\textit{(Right)}. The sub-mm datapoint can be approximately fit when increasing the optical depth 
of the other particles sizes at 855 $\mu m$ as well, showing that there is a linear degeneracy between particle sizes and mass.
 
The attempt to fit the data with the high opacity CPD from sim 1 fails for all attempts of mass/particles combinations. The reason for this is that one always overshoots the near-infrared datapoints, indicating that this envelope is simply too hot to fit the data.

\section{Summary / Discussion}

In this work, we have used data from two radiation hydrodynamic simulations which treat heating and cooling self-consistently (for technical details see \cite{schulik2019} and for the CPD simulations \cite{schulik2020}), in order to generate their plausible spectra via post-processing with the opacities of various reasonable dust populations. The two hydro simulations use two different constant average Rosseland opacities $\kappa_{\rm R} = (1.0,\, 0.01)\,\rm cm^2 \, g^{-1}$, but have otherwise identical parameters, in order to study the structure formation of gaseous envelopes around Jupiter-mass planets (see Table \ref{tab:sim_data}).  {We note that while early mass determination \citep{Keppler2018} indicated planet masses for PDS 70b as $m_b \approx 5$-$8 m_{\rm Jup}$, astrometry-based mass determination \citep{wang2020} moves the planetary masses lower, towards $m_b \approx 2$-$4 m_{\rm Jup}$ and $m_c \approx 1$-$3 m_{\rm Jup}$ for PDS 70c.} 

Due to the nature of our flux-limited diffusion approach, which was used to generate the hydrodynamic data, it was not possible to obtain spectral information directly from those simulations. Therefore we chose to match the average optical depths found in the hydro simulations with rescaled variants of the dust properties known from the studies of the ISM \citep{draine2003, compiegne2011} and meteoritic as well as cometary material \citep{scott2007, brownlee2014, guttler2019}. Motivated by those, we scale the ISM distribution up so that the maximum particle sizes are between 1 $\mu$m and 1 mm, with the minimum sizes given by the corresponding \cite{draine2003}-distribution.

From the dust properties we computed the wavelength resolved opacity functions $\kappa(\lambda)$,  {which are} necessary to generate spectra and synthetic imaging. The geometric optics region of $\kappa(\lambda)$, where those functions are  {near-}constant, was then used to match the sum of all mass-weighted Rosseland mean opacities $\kappa_{\rm R}$ in order to be as self-consistent as possible. 

We used the $\kappa_{\rm R} = 1.0\,\rm cm^2 \, g^{-1}$ simulation from \cite{schulik2020}, which is similar in structure to the outcome of previous work \citep{ayliffebate2009,  judith2017, lambrechts2019}, to introduce key concepts and explain when CPD contributions to the spectrum are important. However, as our results in our previous work indicated that mainly $\kappa_{\rm R} = \rm 0.01\, cm^2 \, g^{-1}$ leads to efficient formation of a CPD, spectral data based on this simulation  { is used for the remainder of the work that we present here. }


Concerning the detectability of continuum features we note how the growth already from 1 $\mu$m to 10 $\mu$m in particle size flattens the particle spectra significantly, see Fig \ref{fig:opacities}. Water nominally should still exhibit features, but those are obscured by the grey carbonates.
Carbonates are furthermore important for the extinction in our models, as already noted in \cite{ossenkopf1994}. The main remaining feature that models with 10 $\mu$m sized particles exhibit is the olivine emission feature. For mm-sized particles the NIR and MIR regions are featureless, but clearly broader than a single black-body.

Our modeling generally shows that the spectral continuum background in the wavelength range of 30-500 $\mu$m is a good hunting ground for CPD contributions. Further details would need to be elucidated on a case-by-case basis. 
The NIR shows generally planet dominated emissions, tentatively validating the approach by \cite{haffert2019}, but cautioning and noting that due to our limited gravitational potential in the hydro simulations, we are likely overestimating the planetary contribution. Therefore the L band could in reality be significantly more dominated by the CPD signal, which was also concluded in \cite{judith2019}.

Our key finding is that the greyness of mm-sized particles translates into a black-body like FIR-slope of $\lambda F_{\lambda} \sim \lambda^{-3}$. 
While one could expect to find shallower slopes from 2D-optically thin disc models \citep{backman1992} or 3D effects,  {those do not show up in our simulations}.
This is to be contrasted with the FIR-slopes of $1-10$ $\mu$m particles which should give $\lambda F_{\lambda} \sim \lambda^{-5}-\lambda^{-4}$. With the preceeding discussion of low opacity for small particles in the FIR due to the scattering slope, the $\lambda F_{\lambda} \sim \lambda^{-3}$ slope should be detectable even in the presence of smaller particles according to an MRN distribution and thus give a clear signal for dust growth.

It is well-known, however, that realistic icy coatings, and crystalline structures can reduce the opacity slope to $\kappa_{\rm R}\sim \lambda^{-1}$ \citep{ossenkopf1994, jaeger1998}. Additionally fluffy growth and warming effects from 20 to 60 K change the FIR slopes of olivines through activation of phonon modes \citep{henning1997, mennella1998, demyk2017}. It is however unlikely that all those effects can conspire to produce an entirely grey opacity function in the FIR. Studies like those performed for pre-stellar cores \citep[e.g.][]{chacon2019}, where a change in the submm extinction slope can be detected in one object, should be performed with ALMA  and would be able to put a final word on whether grey, and hence mm-sized particles exist in objects like PDS 70c.

Dust continuum modelling is a particularly strong tool to get overall spectral shapes correct, as opposed to, e.g., the modeling of individual gas lines.
The infrared-data from \cite{haffert2019} and the submm-data from \cite{isella2019} allowed us to attempt exactly this.
The 'late'-type models which were designed to fit the derived dust masses in \cite{isella2019}, move the CPD spectrum into a reasonable range between the data points. As we used a factor 3 decrease in effective dust-to-gas ratios to match the overall optical depth of the hydro runs with the dust populations, this means that our preferred dust mass for the PDS 70c disc is $1/3$ below the previously reported value. With the modeling performed, it seems that only using mm-sized particles results in a spectrum that is broad enough to fit this data and consistent with a planetary body and at a similar mass. The residual planetary black-body fit resulting from this is  {excessively large} ($R_{\rm P}\approx (4.5 \pm 0.5) R_{\rm Jup}$ as this  {is an unrealistic planet size} \cite{mordasini2017} at the PDS 70 age of 3-5 Myrs \citep{Keppler2018}). We note however that increasing the dust mass arbitrarily, we can fit PDS 70c also with smaller dust sizes, but this requires the CPD to be optically thick.

This can be taken as an indicator for ongoing dust growth in PDS 70c for the following reason. The final width of our CPD spectra will be less wide  {at the NIR end} than the spectra presented here, due to our overestimation of the planetary contribution (the simulated contribution, not the black-body fit of the data). Therefore, more realistic CPD simulations will require grown particles to fit both ends of the spectrum simultaneously. 
 While our overall fit seems to indicate that the dust mass estimate for PDS 70c is correct, we caution that the 'late' type sims are not fully self-consistent simulations. Realistic CPDs of this mass might be hotter, but unfortunately is is unfeasible at the moment to construct such CPDs in simulations through successive accretion of gas, due to prohibitively long simulation timescales.  Therefore, as the radiation hydrodynamic models need some improvement, so do we need more data points in order to ascertain the uniqueness of the fit.

The interpretation of what  {real age the} 'late'  {runs correspond to} for an object like PDS 70c is unclear. While in our simulations, the CPDs have only formed and started to accrete, a PDS 70c-like mass would be a late stage of evolution, while in the interpretaion of \cite{isella2019}, a PDS 70c-like mass would be late because they expected more massive objects and interpret the majority of the mass as being already accreted into the planet. This  {ambiguity} is obviously not settled and awaits future work.

 {Beyond this, \citet{shibaike2024} recently inferred in 1-D in-situ dust growth models that dust growth to 1 mm size is plausible in the dense CPD environment, and \citet{karlin2023} showed in multi-fluid dusty 3-D simulations that direct delivery of dust of up to $100 \mu$m in size to the CPD is realistic. While the latter simulations, as our earlier work in \citet{schulik2020}, were limited in their radial extent surrounding the planet due to the 3-D approach, the 2.5-D radiation hydrodynamic simulations of \cite{marleau2023} resolved the planet. They show that the circumstellar midplane gas flow, which should carry the largest particles, can be distributed over large radii in the circumplanetary midplane. While all those results were obtained in different setups and under different assumptions, and a fully self-consistent 3-D radiation hydrodynamic multi-fluid dust-growth solution of the problem is yet lacking, a plausible synthesis seems to be that mm-sized dust can either grow or be delivered to the CPD. The remaining outstanding question is whether the small dust can be removed quickly enough either through growth or drift to allow for the spectral distribution consistent with the mean optical depths assumed throughout this work.}
 {Nevertheless, our study shows that the spectral width from 3$\mu$m-$1$mm for an object like the protoplanetary disc surrounding PDS 70c can be used to infer active dust growth, a necessary step in the formation of satellites.}

\begin{acknowledgements}
 MS wants to thank Sebastiaan Haffert and Per Bjerkeli for discussing flux conversion at various wavelengths and Sebastian Lorek for extensive discussion on dust populations in the solar system. MS was supported by a project grant from the Swedish Research Council (grant number 2014-5775). AJ wants to thank the support by the Knut and Alice Wallenberg Foundation (grant number 2017.0287), the European Research Council (ERC Consolidator Grant 724687-PLANETESYS) and VR grant 2018-04867. ML thanks the Knut and Alice Wallenberg Foundation under grant 2017.0287.
 All the simulations presented in this work were performed on resources provided by the Swedish National Infrastructure for Computing (SNIC) at Lunarc in Lund University, Sweden, and the entire team is grateful for being supported with their expertise.
\end{acknowledgements}

\bibliographystyle{aa}
\bibliography{accretion} 

\begin{appendix} 

\section{Conversion of observational data}
\label{sec:appendix_dataconversion}

The comparison of our simulated spectral data with the real observations of PDS 70c requires some conversion operations, which are described in the following.
Radmc3d puts out spectral data $F_{\rm \nu}$ in $\rm erg\, s^{-1}\,cm^{-2}\,Hz^{-1}$ at 1 pc distance. The conversion of those fluxes to the universal quantity $\nu F_{\nu} = \lambda F_{\lambda}$ is trivial via $F_{\rm \lambda} = F_{\rm \nu}\;\times \lambda^2/c$, and then rescaled with the distance of PDS70 (113 pc).

The infrared data is given as magnitude differences $\Delta m_{B}$ for band B w.r.t the host star \citep{Keppler2018, muller2018, haffert2019} having magnitudes $m_{\rm PDS70, B}$ and the stellar magnitudes are listed in \cite{Keppler2018}. Conversion to fluxes is then performed via the fundamental relation

\begin{align}
F_{\rm B} = F_{\rm 0, B} \times \bar{\lambda} \times 10^{-(\delta m_{\rm B} + m_{\rm PDS70, B}) /2.5}    
\end{align}

where $\bar{\lambda}$ is the band-midpoint, and $F_{\rm 0, B}$ are the standardized flux-zeropoints for Vega, and therefore we use $F_{\rm 0, Ks} = 4.3 \times 10^{-7}\rm erg/cm^2/\mu m$,  $F_{\rm 0, L'} = 5.26 \times 10^{-8}\rm erg/cm^2/\mu m$. While those are not the most precise values (e.g. for NACO/SPHERE a more involved calibration process is described in \cite{vigan2016}), they reproduce the values in \cite{muller2018} for PDS 70b, and the aim of our fit is the broadness of the spectrum. 
Then, with $m_{\rm Ks} = 7.75$, $m_{\rm L'} = 8.84$, $\delta m_{\rm Ks} = 8.8$, $\delta m_{\rm L'} = 6.6$, and the previous specifications, we get $\lambda F_{\lambda} (2.18 \mu m) = 1.1 \times 10^{-13} \rm erg\, s^{-1}\,cm^{-2}$,  $\lambda F_{\lambda} (3.8 \mu m) = 1.1\times 10^{-13} \rm erg\, s^{-1}\,cm^{-2}$.

The sub-mm data from \cite{isella2019} is taken as $0.1\rm mJy/beam$, which is mentioned as the peak value, and we estimate 1 beamwidth as setting most of the flux. Hence $\nu F_{\nu}(855 \mu m) = \frac{c}{855\mu m} 10^{-27}\rm erg\, s^{-1}\,cm^{-2}\,Hz^{-1} = 3.5 \times 10^{-16}\rm erg\, s^{-1}\,cm^{-2}$. The conversion factor from $\rm erg\, s^{-1}\,cm^{-2}$ to $\rm W\,m^{-2}$ is $10^{-3}$. 

Error bars for the Ks, Lp bands are taken from the literature, the error bar for the submm flux has been estimated the 2-sigma width of the data, assuming the CPD signal is a Gaussian.

\end{appendix} 


%
%

\end{document}